\documentclass[pra,reprint,superscriptaddress,floatfix]{revtex4-1}

\usepackage{graphicx}   
\usepackage{amsmath}    
\usepackage{amssymb}    
\usepackage{braket}     
\usepackage{bm}         
\usepackage{tikz}       
\usetikzlibrary{decorations.markings} 
\usepackage{comment}
\usepackage{etoolbox}
\usepackage{hyperref}
\hypersetup{
    colorlinks=true,
    linkcolor=blue,
    citecolor=black,
    urlcolor=black
}
\usepackage{cleveref}
\usepackage{url}

\usepackage{xcolor}
\usepackage[most]{tcolorbox}

\newif\ifhighlight
\highlightfalse

\DeclareRobustCommand{\highlighttext}[1]{%
  \ifhighlight
    {\color{red}#1}%
  \else
    #1%
  \fi
}

\newtcolorbox{mybox}[1][]{%
    enhanced,                     
    boxrule=0.5mm,               
    sharp corners,               
    breakable,                   
    boxsep=5pt,                  
    title=#1,                    
    fonttitle=\bfseries,         
    boxed title style={%
        colback=white,           
        colframe=black,          
        boxrule=0.5mm,           
        sharp corners,           
        interior hidden,         
    },
    overlay unbroken={
        \draw[black, line width=0.5mm] 
            (frame.south west) ++(0,0) -- 
            (frame.south east) ++(0,0);
    },
}

\crefname{section}{Section}{Sections}
\Crefname{section}{Section}{Sections}
\crefname{equation}{Eq.}{Eqs.}
\Crefname{equation}{Equation}{Equations}

\newcommand{\Tr}{\operatorname{Tr}}

\begin{document}

\title{Equivalence between the second order steady state for the spin-Boson model and its quantum mean force Gibbs state}

\author{Prem Kumar}
\email{premkr@imsc.res.in}
\affiliation{Optics and Quantum Information Group, The Institute of Mathematical Sciences, C.I.T. Campus, Taramani, Chennai 600113, India.}
\affiliation{Homi Bhabha National Institute, Training School Complex, Anushakti Nagar, Mumbai 400094, India}

\author{K. P. Athulya}
\email{athulyakp@imsc.res.in}
\affiliation{Optics and Quantum Information Group, The Institute of Mathematical Sciences, C.I.T. Campus, Taramani, Chennai 600113, India.}
\affiliation{Homi Bhabha National Institute, Training School Complex, Anushakti Nagar, Mumbai 400094, India}

\author{Sibasish Ghosh}
\email{sibasish@imsc.res.in}
\affiliation{Optics and Quantum Information Group, The Institute of Mathematical Sciences, C.I.T. Campus, Taramani, Chennai 600113, India.}
\affiliation{Homi Bhabha National Institute, Training School Complex, Anushakti Nagar, Mumbai 400094, India}

\begin{abstract}
When the coupling of a quantum system to its environment is non-negligible, its steady state is known to deviate from the textbook Gibbs state. The Bloch-Redfield quantum master equation, one of the most widely adopted equations to solve the open quantum dynamics, cannot predict all the deviations of the steady state of a quantum system from the Gibbs state. In this paper, for a generic spin-boson model, we use a higher-order quantum master equation (in system environment coupling strength) to analytically calculate all the deviations of the steady state of the quantum system up to second order in the coupling strength. We also show that this steady state is exactly identical to the corresponding generalized Gibbs state, the so-called quantum mean force Gibbs state, at arbitrary temperature. All these calculations are highly general, making them immediately applicable to a wide class of systems well modeled by the spin-Boson model, spanning a diverse range of topics, from nanomaterials to various condensed-phase processes, and quantum computing (for e.g., environment induced corrections to the steady state of a superconducting qubit). As an example, we use our results to study the dynamics and the steady state of a solid-state double-quantum-dot system under physically relevant choices of parameters.
\end{abstract}

\maketitle

\section{Introduction}
The Bloch-Redfield (BR) master equation (ME) is a cornerstone in the theory of open quantum systems and is widely adopted in diverse fields such as quantum chemistry and biology, thermodynamics, quantum information, computing, and technology to model the effect of an environment on a quantum system \cite{gardiner2004quantum, breuer2002theory, weiss2012quantum}. It is known that the textbook Gibbs state is the steady state (SS) of a quantum system only in the vanishing system-environment (SE) coupling limit. The BR-ME correctly predicts the $O(\lambda^2)$ deviation of the coherence of a quantum state from the Gibbs state \highlighttext{\cite{PhysRevB.71.035318, thingna2012generalized, PhysRevLett.121.070401}}, where $\lambda$ is the SE coupling parameter. However, the \highlighttext{populations} predicted by BR-ME is correct only up to $O(1)$. In fact, it is a general fact that a perturbative ME up to $O(\lambda^{2n})$ is able to predict the coherence of SS and populations up to $O(\lambda^{2n})$ and $O(\lambda^{2n-2})$, respectively \cite{thingna2012generalized}. Since the BR-ME is a $O(\lambda^2)$ ME, to obtain a more accurate dynamics of an open quantum system and to capture any deviation of the SS populations from the Gibbs state, we need to go to a higher order ME.

The time-convolutionless (TCL) projection operator method is a scheme to perturbatively derive a ME of the order of even powers of the SE coupling parameter $\lambda$ \cite{breuer2002theory}. An $O(\lambda^{2n})$ TCL-ME is usually denoted as TCL2n-ME. In fact, BR-ME is equivalent to TCL2.
We hence need to solve TCL4 in order to correctly predict the SS populations up to $O(\lambda^2)$.
But higher order TCL generators are quite difficult to calculate. Before getting into further details about the SS of a quantum state, we discuss a related concept called the mean force Gibbs state (MFGS), the generalized Gibbs state of a quantum system coupled at a non-negligible strength to its environment \cite{trushechkin2022quantum}.

The MFGS is formally defined as
\begin{align}
    \hat{\rho}_{MF} &= \Tr_E Z^{-1} e^{- \beta \hat{H}_{SE}},
\end{align}
where $\hat{H}_{SE}$ is the full SE Hamiltonian, $Z = \Tr e^{-\beta \hat{H}_{SE}}$ and $\beta$ is the inverse temperature of the environment. The MFGS has many applications, for example in the field of strong-coupling quantum thermodynamics \cite{miller2018hamiltonian, seifert2016first, philbin2016thermal, jarzynski2004nonequilibrium, jarzynski2004nonequilibrium, campisi2009fluctuation, rivas2020strong, talkner2020colloquium, strasberg2020measurability}.
Perturbative expressions for the MFGS in weak and strong coupling are known analytically \cite{cresser2021weak, kumar24ultra}. In the ultrastrong regime, it has also been verified to be the SS of the corresponding strong coupling ME \cite{PhysRevA.106.042209}.

It has been proven that there always exists a finite and small value of $\lambda$ for which the SS and MFGS correspond to each other, although the proof of whether this is true for arbitrary values of $\lambda$ is unavailable \cite{trushechkin2022quantum, jaksic1996on, bach2000return, merkli2007decoherence, merkli2008resonance, merkli2008dynamics, merkli2001positive, frohlich2004another, merkli2022dynamics, merkli2022dynamicsb}. In addition, this bound on $\lambda$ is known to decrease with decreasing temperature. In fact, at zero temperature, a violation of the equivalence between SS and MFGS has also recently been reported \cite{crowder2024invalidation}. This makes the proof of the correspondence between the SS and the MFGS up to $O(\lambda^{2n})$ at arbitrary temperature of high interest.

Under the assumption that, in the expression for the SS, the value of the coherences can be analytically continued to the \highlighttext{populations}, Thingna et al. have proven that \highlighttext{second order} SS matches with the corresponding expression for the MFGS, without requiring the calculation of complicated higher order generators for time evolution \cite{thingna2012generalized}. Using a more direct technique, DiVincenzo et al. \cite{PhysRevB.71.035318} have proven this result for a spin-Boson model (SBM) with Ohmic spectral density and exponential cutoff in the large cutoff limit. Recently, TCL4-ME has been used to \highlighttext{numerically} show this result at zero temperature \highlighttext{for ohmic spectral density with exponential cutoff.} \cite{crowder2024invalidation}. In addition, cumulant ME has recently been used to give a better approximation to \highlighttext{the second order} SS than BR and other equations \cite{PhysRevE.110.014144, CummulantMEVsBRME}.

The SBM consists of a two level quantum system coupled to many Bosonic modes, uncoupled from each other \cite{breuer2002theory, weiss2012quantum}. It is a powerful model that has found wide scale applications in diverse fields from quantum chemistry and biology to many body physics and quantum information, computing, and technology. For example, the SBM has been used to study bio-molecule systems like chromophores and photosynthetic complexes \cite{gilmore2006criteria, Gilmore_2005, gilmore2008quantum}, trapped ions \cite{porras2008mesoscopic}, electron transfer reactions \cite{cheche2001dynamics, merkli2013electron} and superconducting circuits \cite{magazzu2018probing}.

Here, for a generic SBM at arbitrary temperature, we analytically prove the equivalence between SS and MFGS up to $O(\lambda^2)$ in coupling using TCL4 generator. We do this in a very general setting, while making just one assumption \highlighttext{that} the spectral density, $J(\omega)$, \highlighttext{ can be analytically continued for negative values of $\omega$, with the property that it is an odd function of $\omega$ (i.e., $J(\omega) = - J(-\omega)$)}. This assumption was made to simplify the complicated analytical calculations. We do note that this assumption is indeed valid for a wide range of situations of physical interest. In particular, we apply our results to a semiconductor double-quantum-dot (DQD) embedded on a substrate \cite{purkayastha2020tunable, PhysRevLett.104.036801, colless2014raman, PhysRevLett.120.097701, PhysRevB.97.035305, PhysRevApplied.9.014030, PhysRevX.6.041027, doi:10.1126/science.aaa2501} which models a charge qubit coupled to a phononic bath \cite{PhysRevA.57.120, BRANDES2005315}.

Since SBM has wide scale applications, this makes our results of widespread interest. Our results are also an important step in the direction of exactly calculating the generators for higher order MEs in a general setting, in order to move to the non-Markovian regime that serves as a better model for many systems of physical interest. Moreover, our results also serve as a validation of the analytic continuation assumption used by Thingna et al. \cite{thingna2012generalized} for SBM \highlighttext{under the assumption that the spectral density is an odd function.}

This article is arranged in the following manner. In \cref{sec_basic_concepts}, we introduce some basic concepts. In \cref{sec_O2_correction}, we provide our analytical results regarding the \highlighttext{second order} SS coherence and \highlighttext{populations} for the generic SBM and show that it exactly matches with the corresponding MFGS. In \cref{sec:DQD}, we apply our results to study the dynamics and \highlighttext{second order} SS of the solid-state DQD system under physically realistic parameter regime. In \cref{sec_conclusion}, we provide conclusions, discussions and future directions.

\section{Basic Concepts}\label{sec_basic_concepts}

\subsection{TCL-ME and its steady state}

For a quantum state $\hat{\rho}(t)$, the TCL-ME can be written in the following form \cite{breuer2002theory}
\begin{align}\label{eqn_TCL}
    \dot{\hat{\rho}}_S(t) &= \sum_{n=0}^\infty \lambda^{2n} \mathcal{F}^{(2n)}(t) [\hat{\rho}(t)].
\end{align}
Here, $\lambda$ is a parameter quantifying the strength of the SE coupling and $\mathcal{F}^{(0)}(t) [\hat{\rho}(t)] = -i [\hat{H}_S, \hat{\rho}(t)]$ is the free Hamiltonian part of the generator of time evolution and likewise $\mathcal{F}^{(2)}(t)$ is the generator for the familiar BR-ME, while terms corresponding to $n>1$ give rise to corrections to the BR-ME. \highlighttext{Throughout this manuscript, we set $\hbar = 1$.}

The \highlighttext{second order} SS of the TCL-ME can then be calculated perturbatively by solving the following equation
\begin{align}\label{eqn_TCl_steady_state_perturbative}
\left( \mathcal{F}^{(0)} + \lambda^2 \mathcal{F}^{(2)} + ... \right) [\hat{\rho}^{(0)} + \lambda^2 \hat{\rho}^{(2)} + ...] = 0.
\end{align}
Here, $\hat{\rho}^{(0)}$ is the Gibbs state and the higher order terms are corrections over it.

\subsection{The spin-Boson model} \label{sec_SBM_model}
For an open quantum system, let the full SE Hamiltonian be given as
\begin{align}
    \hat{H}_{SE} &= \hat{H}_S + \hat{H}_E + \hat{H}_I \label{eqn_H_SE},
\end{align}
where, $\hat{H}_S$ and $\hat{H}_E$ are the system and environment free Hamiltonian, respectively, and
\begin{align}\label{eqn_H_interaction}
    \hat{H}_I &= \lambda \hat{A} \otimes \hat{B}
\end{align}
is the SE interaction operator, where $\hat{A}$ and $\hat{B}$ are system and environment operators, respectively. Although this is not the most general interaction Hamiltonian possible, it is widely studied, and here we restrict our discussion to this for simplification.
Then, the general Hamiltonian for the SBM is given as \cite{breuer2002theory, weiss2012quantum}
\begin{align}\label{eqn_sb_hamiltonian}
    \hat{H}_S &\equiv \frac{\Omega}{2} \hat{\sigma}_3,\\
    \hat{H}_E &= \sum_k\left[\frac{\hat{p}_k^2}{2 m_k}+\frac{1}{2} m_k \omega_k^2 \hat{q}_k^2\right],\\
    \hat{A} &= a_3 \hat{\sigma}_3 - a_1 \hat{\sigma}_1, \label{eqn_spin_boson_A}\\
    \hat{B} &= \sum_k c_k \hat{q}_k, \label{eqn_spin_boson_B}
\end{align}
where $\hat{\sigma}_i$ is the Pauli matrix. Also, $\hat{p}_k$ and $\hat{q}_k$ are the momentum and position operators of the $k$th Bosonic oscillator with mass $m_k$, frequency $\omega_k$ and SE coupling strength $c_k$. Then the spectral density is defined as
\begin{align}
    J(\omega) &= \sum_k \frac{c_k^2}{m_k \omega_k}\delta(\omega - \omega_k).
\end{align}

To simplify the calculations, let us use the vectorized notation for the system state as following,
\begin{align}\label{eqn_vectorization_of_qubit_state}
    v_i(t) &= \Tr \{ \hat{\sigma}_i \hat{\rho}_S(t) \},
\end{align}
where $\hat{\sigma}_0$ is the identity matrix. Then the TCL-ME for this system can be written as \cite{breuer2001time},
\begin{align}
    \dot{\vec{v}}(t) = \sum_{n=0}^\infty \lambda^{2n} F^{(2n)}(t) \vec{v}(t),
\end{align}
where $F^{(2n)}(t)$ is a $4 \times 4$ matrix with matrix elements $F_{mn}^{(2n)}(t)$ which can be obtained from the superoperator $\mathcal{F}(t)$ (\cref{eqn_TCL}) as
\begin{align} \label{eqn_operator_to_matrix}
    F_{mn}^{(2n)}(t) \equiv \Tr\{\hat{\sigma}_m \mathcal{F}^{(2n)}(t) [\hat{\sigma}_n]\}.
\end{align}
In the asymptotic limit, let us define
\begin{align}
    F_{mn} &\equiv \lim_{t \to \infty} F_{mn}(t),\\
    v_{n} &\equiv \lim_{t \to \infty} v_{n}(t).
\end{align}

The coefficients $F_{mn}$ and $v_n$ have a perturbative expansion in $\lambda^2$ as
\begin{align}
    F_{mn} &= F_{mn}^{(0)} + \lambda^2 F_{mn}^{(2)} + \lambda^4 F_{mn}^{(4)} + ...,\\
    v_{n} &= v_{n}^{(0)} + \lambda^2 v_{n}^{(2)} + \lambda^4 v_{n}^{(4)} + ...,
\end{align}
and can systematically be calculated order by order \cite{breuer2001time, breuer2002theory}, as we will show for TCL2 and TCL4.

\section{Second order correction to steady state of spin-Boson model}\label{sec_O2_correction}

\subsection{Correction to coherence} \label{sec_coh_correction}

\highlighttext{In this section, for completeness, we provide the derivation for the second order SS coherences for a SBM and its equivalence with the corresponding MFGS coherences, which has previously been derived elsewhere \cite{PhysRevB.71.035318, thingna2012generalized,PhysRevLett.121.070401, purkayastha2020tunable}}.

In \cref{appendix_general_results}, we show that, \highlighttext{given that $\hat{A}$ is the operator through which the system is coupled to the environment (see \cref{eqn_H_interaction})}, for a system operator $\hat{O}_1$ defined such that
\begin{align}\label{eqn_o1_defn}
    [\hat{O}_1, \hat{A}] \equiv 0,
\end{align}
even at arbitrarily strong SE coupling, we have
\begin{align}\label{eqn_evolution_of_O1}
    \frac{d}{dt} \braket{\hat{O}_1}_{\hat{\rho}_S(t)} = \Tr \{ -i [\hat{O}_1, \hat{H}_S] \hat{\rho}_{S}(t)\}.
\end{align}
That is, the expectation value of $\hat{O}_1$ is evolved by only the free Hamiltonian. Then, in the large time limit, we have
\begin{align}\label{eqn_o1_SS_derivative}
    \frac{d}{dt} \braket{\hat{O}_1}_{\hat{\rho}_S(t)} &= 0 = \lim_{t \to \infty} \Tr \{ -i [\hat{O}_1, \hat{H}_S] \hat{\rho}_{S}(t)\}.
\end{align}
Defining
\begin{align}\label{eqn_O2_SS_0}
    \hat{O}_2 \equiv [\hat{O}_1, \hat{H}_S], 
\end{align}
we then have that,
\begin{align}\label{eqn_o2_exp_zero}
    \braket{\hat{O}_2}_{SS} &= 0.
\end{align}
Here, $\braket{\hat{O}}_{SS}$ denotes \highlighttext{the exact} SS expectation value of the operator $\hat{O}$.

As proven in \cref{appendix_general_results}, we note that this is a non-perturbative result which holds for arbitrary value of $\lambda$ and hence holds not just for TCL2n-ME for arbitrary $n$, but also in intermediate and strong SE coupling regime where TCL fails. This result is also valid for a generic $d$-dimensional system.

This result is relevant in the present discussion because choosing $\hat{O}_1 = \hat{A}$, we get (from \cref{eqn_sb_hamiltonian}, \cref{eqn_spin_boson_A} and \cref{eqn_O2_SS_0}) that
\begin{align}
    \hat{O}_2 = [\hat{A},\hat{H}_S] \propto \hat{\sigma}_2,
\end{align}
which means that under the present SBM, to all orders in TCL, we have
\begin{align}
    \braket{\hat{\sigma}_2}_{SS}^{(2n)} &= v_2^{(2n)} = 0,\\
    \implies v_2^{(2)} &= 0. \label{eqn_v_2_is_zero}
\end{align}
As a side note, we also prove in \cref{appendix_general_results} that the corresponding MFGS expectation value for $\hat{O}_2$ (\cref{eqn_o2_exp_zero}) is also zero at arbitrary SE coupling.

Next, to calculate $v_1^{(2)}$, we extract the $O(\lambda^2)$ term from \cref{eqn_TCl_steady_state_perturbative} as,
\begin{align} \label{eqn_O2_TCL_eqn_coherence}
F^{(0)} \vec{v}^{(2)} + F^{(2)} \vec{v}^{(0)} &= 0.
\end{align}
The zeroth order terms in this equation are given as
\begin{align}
    F^{(0)} &= \left(
    \begin{array}{cccc}
     0 & 0 & 0 & 0 \\
     0 & 0 & -\Omega  & 0 \\
     0 & \Omega  & 0 & 0 \\
     0 & 0 & 0 & 0 \\
    \end{array}
    \right),\\
    \vec{v}^{(0)} &= \left(
\begin{array}{c}
 1 \\
 0 \\
 0 \\
 v_3^{(0)} \\
\end{array}
\right).
\end{align}
The 3rd row of the matrix on the LHS of \cref{eqn_O2_TCL_eqn_coherence} then gives us,
\begin{align}\label{eqn_v12_expression}
    v_1^{(2)} &= -\frac{F_{20}^{(2)} + v_3^{(0)} F_{23}^{(2)}}{\Omega}.
\end{align}

The matrix elements for the TCL2 generator ($F_{20}^{(2)}$ and $F_{23}^{(2)}$) can be evaluated for an arbitrary open quantum system (see \cite{breuer2001time, breuer2002theory}). The details of the calculations and solutions of all the elements of the TCL2 generator are given in \cref{appendix_TCL_2_gen_and_sol}. For the present case, for example, we have
\begin{align}
    F_{20}^{(2)} &= -4 a_1 a_3 \lim_{t \to \infty}\int_0^t dt_1\eta \left(t-t_1\right) (\cos \left(\left(t-t_1\right) \Omega \right)-1),\label{eqn_f2_20_integrand}\\
    F_{23}^{(2)} &= -4 a_1 a_3 \lim_{t \to \infty} \int_0^t dt_1 \nu \left(t-t_1\right) \sin \left(\left(t-t_1\right) \Omega \right).\label{eqn_f2_23_integrand}
\end{align}
Here, $\nu(t)$ and $\eta(t)$ are defined as the real and imaginary part of the environment's two point correlation function (\cref{eqn_spin_boson_B}),
\begin{align}
    \eta(t) &\equiv \text{Im}\{\braket{\hat{B}, \hat{B}(-t)}\},\\
    \nu(t) &\equiv \text{Re}\{\braket{\hat{B}, \hat{B}(-t)}\},
\end{align}
and in terms of the spectral density, can be evaluated to be
\begin{align}
    \eta(t) &= -\int_0^\infty d\omega J(\omega ) \sin (t \omega ),\label{eqn_eta_expr}\\
    \nu(t) &= \int_0^\infty d\omega f(\omega ) \cos (t \omega ), \label{eqn_nut_expr}
\end{align}
where, we have defined
\begin{align}\label{eqn_definition_of_f_omega}
    f(\omega) &\equiv J(\omega) \coth(\beta \omega/2).
\end{align}

For a general spectral density, $J(\omega)$, in \cref{eqn_f2_20_integrand} and \cref{eqn_f2_23_integrand}, we first do the $t_1$ integral analytically, and then take the large $t$ limit. In this case, the result can be evaluated as,
\begin{align}
    F_{20}^{(2)} &= 4 a_1 a_3 \Omega ^2 \int_0^\infty d \omega \frac{J(\omega )}{\omega ^3-\omega  \Omega ^2},\\
    F_{23}^{(2)} &= 4 a_1 a_3 \Omega  \int_0^\infty d \omega  \frac{f(\omega )}{\omega ^2-\Omega ^2},
\end{align}
which, using \cref{eqn_v12_expression}, yields
\begin{align}
    v_1^{(2)} &= 4 a_1 a_3 \int_0^\infty d \omega \frac{J(\omega ) \left(\omega  \coth \left(\frac{\beta  \omega }{2}\right) \tanh \left(\frac{\beta  \Omega }{2}\right)-\Omega \right)}{ \omega ^3- \omega  \Omega ^2}.
\end{align}
This matches with the corresponding MFGS expression \cite{cresser2021weak, PhysRevB.71.035318, thingna2012generalized,PhysRevLett.121.070401, purkayastha2020tunable}. It is also possible to verify that $v_2^{(2)} = 0$ using similar calculations, consistent with the more general proof we have provided.

\subsection{Correction to populations}
We cannot use the $O(\lambda^2)$ terms in the perturbative equation for the SS (\cref{eqn_TCl_steady_state_perturbative}) to get the $O(\lambda^2)$ correction to the populations ($v_3^{(2)}$) because when we solve this $O(\lambda^2)$ equation (\cref{eqn_O2_TCL_eqn_coherence}), the term of interest ($v_3^{(2)}$) disappears from the expression. In fact, it is generically true that, in order to get the $O(\lambda^{2n})$ correction to the populations, one needs to go to $O(\lambda^{2n+2})$ ME \cite{thingna2012generalized}. In this case, the $O(\lambda^4)$ terms in the perturbative equation for the SS (\cref{eqn_TCl_steady_state_perturbative}) give us
\begin{align}\label{eqn_O4_TCL_eqn_diagonal}
    F^{(0)} \vec{v}^{(4)} + F^{(2)} \vec{v}^{(2)} + F^{(4)} \vec{v}^{(0)} &= 0.
\end{align}
From the 4th row of the matrix in the LHS of \cref{eqn_O4_TCL_eqn_diagonal}, we get
\begin{align} \label{eqn_diagonal_pop_expression_o2}
    v_3^{(2)} = -\frac{F_{30}^{(4)} + v_3^{(0)} F_{33}^{(4)} + v_1^{(2)} F_{31}^{(2)}}{F_{33}^{(2)}}.
\end{align}
The expressions for $F_{33}^{(2)}$ and $F_{31}^{(2)}$ can be evaluated as in the last section. The expressions for the integrals for all the TCL2 matrix elements as well as their solutions are provided in \cref{appendix_TCL_2_gen_and_sol}.

\begin{figure}[htbp]
    \centering
    \begin{tikzpicture}[scale=1, decoration={markings, mark=at position 0.5 with {\arrow{>}}}]
        \draw[->] (-3.5,0) -- (3.5,0) node[right] {\(\text{Re}(\omega)\)};
        \draw[->] (0,-1) -- (0,1.5) node[above] {\(\text{Im}(\omega)\)};

        \def\eps{0.2}
        \def\zeropole{0}

        \draw[thick, postaction={decorate}] (-3,0) -- (-2 - \eps,0);

        \draw[thick, postaction={decorate}] (-2 - \eps,0) arc (180:0:\eps);

        \draw[thick, postaction={decorate}] (-2 + \eps,0) -- (\zeropole - \eps,0);

        \draw[thick, postaction={decorate}] (\zeropole - \eps,0) arc (180:0:\eps);

        \draw[thick, postaction={decorate}] (\zeropole + \eps,0) -- (2 - \eps,0);

        \draw[thick, postaction={decorate}] (2 - \eps,0) arc (180:0:\eps);

        \draw[thick, postaction={decorate}] (2 + \eps,0) -- (3,0);

        \fill[blue] (-2,0) circle (2pt);
        \fill[blue] (\zeropole,0) circle (2pt);
        \fill[blue] (2,0) circle (2pt);

        \node[below] at (-2,0) {$-2$};
        \node[below] at (-0.2,0) {$0$};
        \node[below] at (2,0) {$2$};

        \node[above right] at (1,0.5) {$C$};
    \end{tikzpicture}
    \caption{Pictorial representation of the form of the contour to be taken for the $\omega$ integral in the definition of the TCL4 coefficients (\cref{eqn_f334_result} and \cref{eqn_f304_result}), where the contour has to be assumed to go above the poles at the real line (represented here as blue dots).}
    \label{fig_contour_avoiding_poles}
\end{figure}
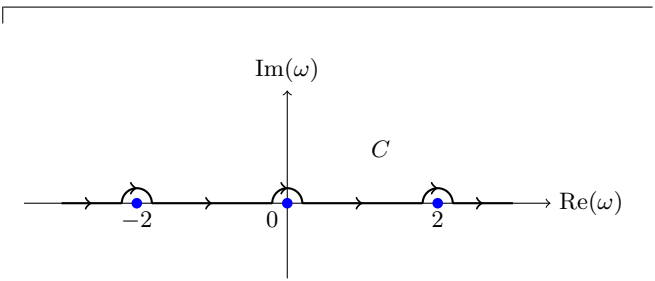

To simplify the integrals for the TCL4 coefficients in \cref{eqn_diagonal_pop_expression_o2}, we assume that $J(\omega)$ can be analytically continued such that we have $J(-\omega) = -J(\omega)$. Then, the TCL4 coefficients can be written as \cite{breuer2001time, TypoAndConventionDiff}
\begin{widetext}
\begin{align}
    F_{33}&^{(4)} = \lim_{t\to \infty} 16 a_1^2  \int_0^{t} dt_1 \int_0^{t_1} dt_2  \int_0^{t_2} dt_3 
    \bigg\{ a_3^2 \bigg[\eta \left(t_1-t_2\right) \eta \left(t-t_3\right) \left(\cos \left(\left(t-t_2\right) \Omega \right)-\cos \left(\left(t-t_1\right) \Omega \right)\right)\notag\\
    &+ \eta \left(t-t_2\right) \eta \left(t_1-t_3\right) \left(\cos \left(\left(t-t_2\right) \Omega \right)-\cos \left(\left(t-t_1\right) \Omega \right)\right)+ \nu \left(t_1-t_2\right) \nu \left(t-t_3\right) \left(\cos \left(\left(t-t_3\right) \Omega \right)-\cos \left(\left(t-t_2\right) \Omega \right)\right)\bigg]\notag\\
    &+ a_1^2 \bigg[\nu \left(t_1-t_2\right) \nu \left(t-t_3\right) \sin \left(\left(t-t_2\right) \Omega \right) \sin \left(\left(t_1-t_3\right) \Omega \right)+\nu \left(t-t_2\right) \nu \left(t_1-t_3\right) \sin \left(\left(t_1-t_2\right) \Omega \right) \sin \left(\left(t-t_3\right) \Omega \right)\bigg]
    \bigg\},
    \label{eqn_a334_triple_integral}
\end{align}
\begin{align}
    F_{30}&^{(4)} = \lim_{t\to \infty} 8 a_1^2  \int_0^{t} dt_1 \int_0^{t_1} dt_2  \int_0^{t_2} dt_3\bigg\{a_1^2 \bigg[\eta \left(t_1-t_2\right) \nu \left(t-t_3\right) \left(\sin \left(\left(t+t_1-t_2-t_3\right) \Omega \right)-\sin \left(\left(t-t_1+t_2-t_3\right) \Omega \right)\right) \notag\\
    &+ 2 \eta \left(t_1-t_3\right) \nu \left(t-t_2\right) \sin \left(\left(t_1-t_2\right) \Omega \right) \cos \left(\left(t-t_3\right) \Omega \right)-2 \eta \left(t-t_3\right) \nu \left(t_1-t_2\right) \sin \left(\left(t_2-t_3\right) \Omega \right) \cos \left(\left(t-t_1\right) \Omega \right)\bigg] \notag\\
    &+ 2 a_3^2 \bigg[\eta \left(t_1-t_3\right) \nu \left(t-t_2\right) \left(\sin \left(\left(t-t_2\right) \Omega \right)-\sin \left(\left(t-t_1\right) \Omega \right)\right)+\eta \left(t-t_3\right) \nu \left(t_1-t_2\right) \left(\sin \left(\left(t-t_2\right) \Omega \right)-\sin \left(\left(t-t_3\right) \Omega \right)\right)\notag\\
    & \quad \quad +\eta \left(t_1-t_2\right) \nu \left(t-t_3\right) \left(\sin \left(\left(t-t_2\right) \Omega \right)-\sin \left(\left(t-t_1\right) \Omega \right)\right)\bigg]
    \bigg\}.\label{eqn_b34_triple_integral}
\end{align}

The TCL4 coefficients are essentially a 5 dimensional integral- three integrals in time variables ($t_1$, $t_2$ and $t_3$) and two in frequency ($\omega_1$ and $\omega_2$, appearing in the functions $\nu(t)$ and $\eta(t)$ each (\cref{eqn_eta_expr} and \cref{eqn_nut_expr})). We proceed by doing the triple time integral first, which can be easily done using a computer algebra system like MATHEMATICA. Afterwards, one has to carefully do the double $\omega$ integrals in a large $t$ limit.

We found (the hard way) that although it does not matter which variable, between $\omega_1$ and $\omega_2$, we choose to integrate first, we must do it consistently for each term in the integrand.
\highlighttext{This is because, although for the full integrand, the result of the integral does not depend upon the order in which it is carried out, this is not true for the individual terms of the integrand. But the discrepancy that arises from these terms essentially cancel out as long as the same choice for the order of the integrals is made for \textit{all} of these terms}.
See \cref{appendix_fubini_effect} for further discussions.

Finally, these integrals are evaluated as,
\begin{align}\label{eqn_f334_result}
    F_{33}^{(4)} &= \frac{1}{2} \int_{-\infty}^\infty d\omega  \bigg\{ \frac{4 \pi  a_1^4 \Omega  f(\omega )}{\left(\omega ^2-\Omega ^2\right)^2} \left[\left(\omega ^2-\Omega ^2\right) f'(\Omega )+2 \Omega  f(\Omega )\right] + \frac{2 \pi  a_1^2 a_3^2}{\left(\omega ^3-\omega  \Omega ^2\right)^2} \notag\\
    & \times \bigg[f(\omega ) \left(4 f(\Omega ) \left(\omega ^2-\Omega ^2\right)^2-2 \omega  \Omega  (\omega +\Omega )^2 f(\omega -\Omega )+\omega  (\omega -\Omega ) (2 \Omega  (\omega -\Omega ) f(\omega +\Omega )-4 f(0) \omega  (\omega +\Omega ))\right)\notag\\
    &\quad \quad - 2 \Omega  J(\omega ) \left(\omega ^2-\Omega ^2\right) ((\omega +\Omega ) J(\omega -\Omega )+(\Omega -\omega ) J(\omega +\Omega ))\bigg]
    \bigg\},
\end{align}
\begin{align}\label{eqn_f304_result}
   F_{30}^{(4)} &= \frac{1}{2} \int_{-\infty}^\infty d\omega \bigg\{ \frac{4 \pi  a_1^4 }{\left(\omega ^2-\Omega ^2\right)^2}\left[f(\omega ) \left(\Omega  \left(\omega ^2-\Omega ^2\right) J'(\Omega )+J(\Omega ) \left(\omega ^2+3 \Omega ^2\right)\right)-2 \omega  \Omega  f(\Omega ) J(\omega )\right] + \frac{4 \pi  a_1^2 a_3^2}{\left(\omega ^3-\omega  \Omega ^2\right)^2} \notag \\
   & \times \left[2 f(0) \omega  \Omega  J(\omega ) \left(\Omega ^2-\omega ^2\right)+f(\omega ) \left((\omega -\Omega )^2 \left(2 J(\Omega ) (\omega +\Omega )^2-\Omega ^2 J(\omega +\Omega )\right)+\Omega ^2 (\omega +\Omega )^2 J(\omega -\Omega )\right)\right]
   \bigg\},
\end{align}
where, the contour for these integrals (\cref{eqn_f334_result} and \cref{eqn_f304_result}) are defined to go above any pole on the real line (see \cref{fig_contour_avoiding_poles}), hence not including the contribution of the poles at the real line and also avoiding the corresponding apparent divergences (See \cref{appendix_TCL4_integral} for details). Using \cref{eqn_diagonal_pop_expression_o2}, we then have the expression of $v_3^{(2)}$ as
\begin{align}
    v_3^{(2)} = -2 a_1^2 \tanh \left(\frac{\beta  \Omega }{2}\right)
    \int_0^\infty &d\omega \frac{J(\omega )}{\left(\omega ^2-\Omega ^2\right)^2}   \left(2 \omega  \Omega  \coth \left(\frac{\beta  \Omega }{2}\right)-\coth \left(\frac{\beta  \omega }{2}\right) \left(\beta  \Omega  (\omega^2 -\Omega^2 ) \text{csch}(\beta  \Omega )+\omega ^2+\Omega ^2\right)\right).
\end{align}
\end{widetext}
This matches with the MFGS value of $v_3^{(2)}$ \cite{cresser2021weak}. We have hence analytically shown that for the generic SBM, all second order SS \highlighttext{components} match with the corresponding MFGS corrections.
\highlighttext{See the Supplemental Material \cite{SupplementaryMaterial} for all these results in electronic form.}

In \cref{appendix_TCL2_and_TCL4_for_Drude_cutoff}, we also provide the values of the TCL2 and TCL4 generators for a special choice of spectral density well studied in literature, namely the Ohmic spectral density with Drude cutoff. For this spectral density, the expressions for the two-point correlation functions $\nu(t)$ and $\eta(t)$ is known in closed form, hence assisting in the evaluation of the TCL4 generator matrix elements. The calculations proceed differently, but we have verified that the results are consistent with our main, more general results, provided here.

\highlighttext{Here, we note how our results presented here differ from those of Crowder et al. \cite{crowder2024invalidation}.
The expression for the asymptotic TCL4 generator obtained by Crowder et al. still has an infinite time integral, which is not the case for our result.
Consequently, they have numerically demonstrated the equivalence between the second order SS and MFGS at zero temperature, whereas our result is analytical one and is valid at arbitrary temperature.
Also, their results were derived under the assumption that the spectral density is of the form $J(\omega) = C \omega^s \exp(-\omega/\omega_c) \Theta(\omega)$ (here $C,s>0$ are some constant and $\Theta(\omega)$ is Heaviside step-function).
Whereas our results are valid for all odd spectral densities.
We do note that their work is valid for a generic $N$ dimensional system, whereas we have restricted ourselves to the generic SBM.}

\section{\label{sec:DQD} Application: Double-Quantum-Dot Model}
The generic SBM Hamiltonian considered by us is known to model a large class of physical systems of interest \cite{purkayastha2020tunable, gilmore2006criteria, Gilmore_2005, gilmore2008quantum, porras2008mesoscopic, cheche2001dynamics, merkli2013electron, magazzu2018probing}. Here, we study the application of the SBM to the solid-state double-quantum-dot (DQD), where we reinterpret the model parameters of the generic SBM (\cref{eqn_sb_hamiltonian} - \cref{eqn_spin_boson_B}) as \cite{purkayastha2020tunable}
\begin{align}
    a_{1} &= \frac{\epsilon}{\Omega},\\
    a_{3} &= \frac{2 t_{c}}{\Omega},\\
    \Omega^2 &= \epsilon^2 + 4 t_c^2.
\end{align}
Here, the parameters $\epsilon$ and $t_{c}$ correspond to the \highlighttext{detuning} and inter dot tunneling \highlighttext{parameters}, respectively.
We calculate the second-order SS (for both coherence and populations) of the solid-state DQD model using the TCL2 and TCL4 generators for the following physically motivated spectral density \cite{purkayastha2020tunable}:
\begin{align}\label{eqn_sinc_spectral_density}
  J(\omega) = \gamma \omega \, \left[1-\text{sinc}\left(\frac{\omega}{\omega_c}\right) \right]\, \exp\left\{-\frac{\omega^2}{2\: \omega_{\text{max}}^2}\right\},
\end{align}
where, we have $\text{sinc}(x) \equiv \sin(x)/x$. This spectral density provides an accurate description of bulk acoustic phonons in GaAs DQDs\cite{PhysRevLett.104.036801, colless2014raman, PhysRevB.97.035305, purkayastha2020tunable}. Here the frequency $\omega_{\text{max}}$ denotes the upper cut-off frequency, while $\omega_{c} = c_{s}/d$, with $c_{s}$ being the speed of sound in the substrate and 
$d$ is the inter-dot distance. The quantity $\lambda^2 \gamma$ governs the coupling strength between the quantum dots and the phonon bath. Also, we note that since this choice of $J(\omega)$ happens to be antisymmetric in $\omega$, our TCL4 results are directly applicable here.
\begin{figure}[htbp]
	\begin{center}
	\resizebox{7.cm}{4.6 cm}{\includegraphics{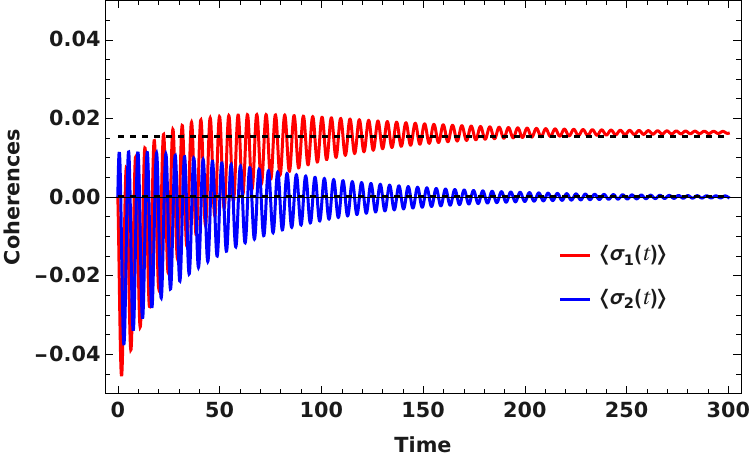}}
   \caption{Coherences of a solid-state DQD system are plotted using TCL2-ME as a function of time for a non-zero detuning. The solid red and blue lines represent the coherences ($\langle\hat{\sigma}_{1}(t)\rangle$ and $\langle\hat{\sigma}_{2}(t)\rangle$, respectively), and are plotted for an initial state with $\langle\hat{\sigma}_{1}(0)\rangle = 0 $, $\langle\hat{\sigma}_{2}(0)\rangle = 0 $,  and $\langle\hat{\sigma}_{3}(0)\rangle = 0 $. The parameters are set as  $\epsilon = 1$, $t_c = 0.5$, $\lambda^2 \gamma = 1.44 \times 10^{-2} $, $\omega_{max} = 8$ and $\beta =1$. The black dashed line represents the MFGS value. Here we observe steady-state coherence for $\hat{\sigma}_{1}(t)$. All energies are measured in the units of $\omega_c = 1$.}
	\label{fig1}
   \end{center} 
\end{figure}
\begin{figure}[htbp]
	\begin{center}
	\resizebox{7.cm}{4.6 cm}{\includegraphics{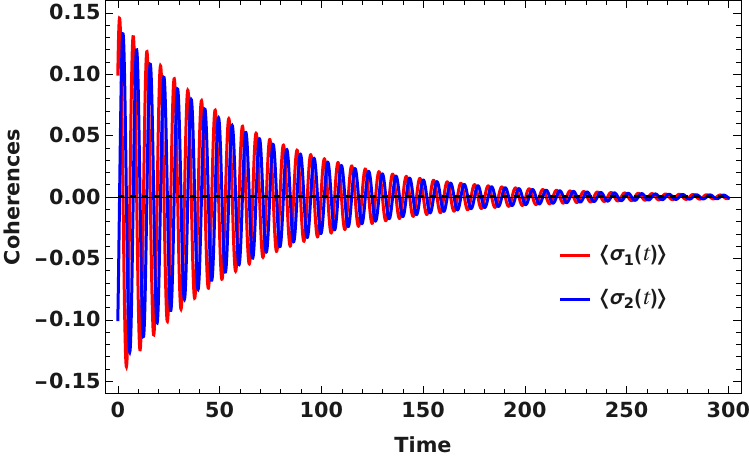}}
   \caption{Coherences of a solid-state DQD system are plotted using TCL2-ME as a function of time for zero \highlighttext{detuning}. The solid red and blue lines represent the coherence ($\langle\hat{\sigma}_{1}(t)\rangle$ and $\langle\hat{\sigma}_{2}(t)\rangle$, respectively), and are plotted for an initial state with $\langle\hat{\sigma}_{1}(0)\rangle = 0.1 $, $\langle\hat{\sigma}_{2}(0)\rangle = -0.1 $,  and $\langle\hat{\sigma}_{3}(0)\rangle = 0 $. The parameters are set as  $\epsilon = 0$, $t_c = 0.5$, $\lambda^2 \gamma = 1.44 \times 10^{-2} $, $\omega_{max} = 8$, $\beta =1$,  and $\omega_c = 1$.}
	\label{fig2}
   \end{center} 
\end{figure}

\begin{figure}[htbp]
	\begin{center}
	\resizebox{7.cm}{4.6 cm}{\includegraphics{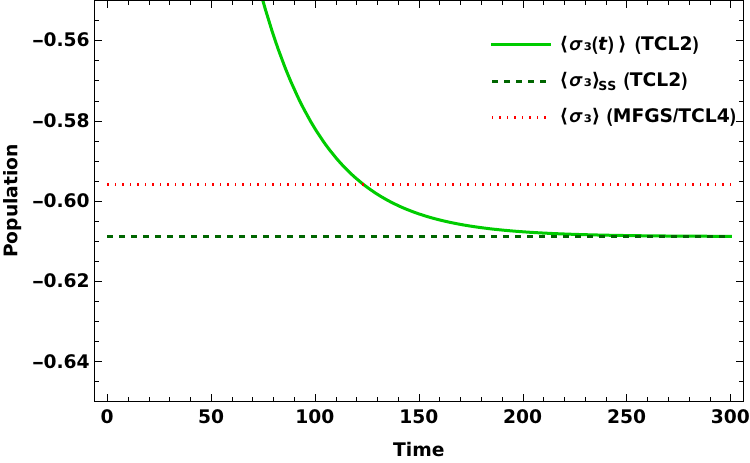}}
   \caption{The \highlighttext{populations} of a solid-state DQD system ($\braket{\hat{\sigma}_3(t)}$, solid green line) is plotted using TCL2-ME as a function of time. In the large time limit, it converges to the \highlighttext{second order} SS value predicted by TCL2 (dashed green line). The dotted red line corresponds the $O(\lambda^2)$ value of $\braket{\hat{\sigma}_3}_{SS}$ derived here using \highlighttext{MFGS}/TCL4. We observe that there is a small correction \highlighttext{to $\braket{\sigma_3}_{SS}$ obtained from TCL2 vs the corresponding MFGS/TCL4 result}. All model parameters here are the same as in \cref{fig1}.}
	\label{fig3}
   \end{center} 
\end{figure}

The expression for the TCL2 generator for a generic spectral density, $J(\omega)$ and its solution, is provided in \cref{appendix_TCL_2_gen_and_sol}. We then evaluate this generator for the present choice of the spectral density (\cref{eqn_sinc_spectral_density}) in order to calculate the dynamics of the solid-state DQD under TCL2.
In \cref{fig1} and \cref{fig2} we plot coherences $\langle\hat{\sigma}_{1}(t)\rangle $  and $\langle\hat{\sigma}_{2}(t)\rangle $ (solid red and blue lines, respectively) as a function of time for two different detuning energies. The black dashed line indicates the coherence for the corresponding MFGS. In both the figures, we observe agreement between the coherences for the \highlighttext{second order} SS and MFGS. As noted in the reference \cite{PhysRevLett.121.070401, purkayastha2020tunable}, in \cref{fig1} we observe a non-zero \highlighttext{second order} SS coherence for  $\langle\hat{\sigma}_{1}(t)\rangle $ for a non-zero \highlighttext{detuning}, while for zero \highlighttext{detuning} (\cref{fig2}), it approaches zero. For $\langle\hat{\sigma}_{2}(t)\rangle $, the \highlighttext{second order} SS coherence approaches zero in both \cref{fig1} and \cref{fig2}, which also agrees with the general argument provided in \cref{sec_coh_correction}.

In \cref{fig3}, the expectation value $\langle\hat{\sigma}_{3}(t)\rangle$, calculated using the TCL2 generator, is shown with a solid green line, while the red dotted line represents the corresponding SS value, correct up to second order, calculated using TCL4. The discrepancy between these two lines is clearly evident in \cref{fig3}, demonstrating that the SS populations obtained from TCL2 differs from the actual $O(\lambda^2)$ value. This discrepancy highlights the necessity of using TCL4 to obtain the SS populations correct to up to $O(\lambda^2)$ for such systems.

\section{Conclusion} \label{sec_conclusion}
In this work, we analytically proved the equivalence between the SS and MFGS of a quantum system for a generic SBM up to $O(\lambda^2)$ in the SE coupling parameter $\lambda$. We achieved this by analytically calculating the relevant elements of the TCL4 generator matrix. We derived this result under the sole assumption that the spectral density, $J(\omega)$, can be analytically continued as an odd function. Although restrictive, we note that this assumption is valid for a wide range of spectral densities of physical interest, well studied in the literature \cite{purkayastha2020tunable}. This work therefore makes it possible to calculate the \highlighttext{populations} corrections for the \highlighttext{second order} SS for a wide variety of problems that are well modeled by the SBM \cite{purkayastha2020tunable, gilmore2006criteria, Gilmore_2005, gilmore2008quantum, porras2008mesoscopic, cheche2001dynamics, merkli2013electron, magazzu2018probing}. For example, in any implementation of quantum computing, e.g., based on superconductors, solid-state DQDs, trapped ion etc., if we model the corresponding qubit-environment interaction under the very generic setting assumed in this work, then our results can be used to \highlighttext{not just} calculate the environment induced second order coherences but also the \highlighttext{populations} in the \highlighttext{second order} SS of the corresponding qubit.

We apply our results to study the dynamics and the SS for a solid-state DQD system under physically realistic model parameters, which has previously been studied in the literature and for which the physically relevant spectral density shares the property assumed by us \cite{purkayastha2020tunable}.

Calculating the TCL4 generator is quite a difficult task. Our results are an important step in the direction of analytically calculating the generators for higher-order MEs in a high level of generality in order to move to the non-Markovian regimes, which serves as a better model for many systems of physical interest.

As a future direction, it will be interesting to relax the assumption of odd spectral density that we made to simplify our calculations for a generic SBM. Another obvious generalization of this result could be moving beyond the SBM to a more general quantum system and a more general SE coupling.

Currently, it is unclear at what values of $\lambda$ and temperature $T$ does the SS of a quantum system correspond to its MFGS \cite{trushechkin2022quantum, jaksic1996on, bach2000return, merkli2007decoherence, merkli2008resonance, merkli2008dynamics, merkli2001positive, frohlich2004another, merkli2022dynamics, merkli2022dynamicsb}. The result is known to hold at a finite but small $\lambda$, but the result for the general value of $\lambda$ is unclear. Also, as we decrease $T$, the bound on $\lambda$ is also found to decrease. In fact, Crowder et al. have reported a deviation of the SS from the MFGS at $O(\lambda^4)$ for $T=0$ \cite{crowder2024invalidation}. It would be interesting to study this phenomenon at finite temperature with the help of analytical expressions for a generic TCL4 generator for the SBM.

\section{Acknowledgment}
The authors would like to thank V. Ranjith for initial discussions related to the Bloch-Redfield equation, its complete-positivity issue, and associated areas.

\bibliographystyle{IEEEtran}
\bibliography{./citation}

\begin{thebibliography}{10}
\providecommand{\url}[1]{#1}
\csname url@samestyle\endcsname
\providecommand{\newblock}{\relax}
\providecommand{\bibinfo}[2]{#2}
\providecommand{\BIBentrySTDinterwordspacing}{\spaceskip=0pt\relax}
\providecommand{\BIBentryALTinterwordstretchfactor}{4}
\providecommand{\BIBentryALTinterwordspacing}{\spaceskip=\fontdimen2\font plus
\BIBentryALTinterwordstretchfactor\fontdimen3\font minus \fontdimen4\font\relax}
\providecommand{\BIBforeignlanguage}[2]{{%
\expandafter\ifx\csname l@#1\endcsname\relax
\typeout{** WARNING: IEEEtran.bst: No hyphenation pattern has been}%
\typeout{** loaded for the language `#1'. Using the pattern for}%
\typeout{** the default language instead.}%
\else
\language=\csname l@#1\endcsname
\fi
#2}}
\providecommand{\BIBdecl}{\relax}
\BIBdecl

\bibitem{gardiner2004quantum}
C.~Gardiner and P.~Zoller, \emph{Quantum noise: a handbook of Markovian and non-Markovian quantum stochastic methods with applications to quantum optics}.\hskip 1em plus 0.5em minus 0.4em\relax Springer Science \& Business Media, Berlin, 2004.

\bibitem{breuer2002theory}
H.-P. Breuer and F.~Petruccione, \emph{The theory of open quantum systems}.\hskip 1em plus 0.5em minus 0.4em\relax Oxford University Press, USA, 2002.

\bibitem{weiss2012quantum}
U.~Weiss, \emph{Quantum dissipative systems}.\hskip 1em plus 0.5em minus 0.4em\relax World Scientific, 2012.

\bibitem{PhysRevB.71.035318}
\BIBentryALTinterwordspacing
D.~P. DiVincenzo and D.~Loss, ``Rigorous born approximation and beyond for the spin-boson model,'' \emph{Phys. Rev. B}, vol.~71, p. 035318, Jan 2005. [Online]. Available: \url{https://link.aps.org/doi/10.1103/PhysRevB.71.035318}
\BIBentrySTDinterwordspacing

\bibitem{thingna2012generalized}
\BIBentryALTinterwordspacing
J.~Thingna, J.-S. Wang, and P.~Hänggi, ``Generalized gibbs state with modified redfield solution: Exact agreement up to second order,'' \emph{The Journal of Chemical Physics}, vol. 136, no.~19, p. 194110, 05 2012. [Online]. Available: \url{https://doi.org/10.1063/1.4718706}
\BIBentrySTDinterwordspacing

\bibitem{PhysRevLett.121.070401}
\BIBentryALTinterwordspacing
G.~Guarnieri, M.~Kol\'a\ifmmode~\check{r}\else \v{r}\fi{}, and R.~Filip, ``Steady-state coherences by composite system-bath interactions,'' \emph{Phys. Rev. Lett.}, vol. 121, p. 070401, Aug 2018. [Online]. Available: \url{https://link.aps.org/doi/10.1103/PhysRevLett.121.070401}
\BIBentrySTDinterwordspacing

\bibitem{trushechkin2022quantum}
\BIBentryALTinterwordspacing
A.~S. Trushechkin, M.~Merkli, J.~D. Cresser, and J.~Anders, ``Open quantum system dynamics and the mean force gibbs state,'' \emph{AVS Quantum Science}, vol.~4, no.~1, p. 012301, 03 2022. [Online]. Available: \url{https://doi.org/10.1116/5.0073853}
\BIBentrySTDinterwordspacing

\bibitem{miller2018hamiltonian}
\BIBentryALTinterwordspacing
H.~J.~D. Miller, \emph{Hamiltonian of Mean Force for Strongly-Coupled Systems}.\hskip 1em plus 0.5em minus 0.4em\relax Cham: Springer International Publishing, 2018, pp. 531--549. [Online]. Available: \url{https://doi.org/10.1007/978-3-319-99046-0_22}
\BIBentrySTDinterwordspacing

\bibitem{seifert2016first}
\BIBentryALTinterwordspacing
U.~Seifert, ``First and second law of thermodynamics at strong coupling,'' \emph{Phys. Rev. Lett.}, vol. 116, p. 020601, Jan 2016. [Online]. Available: \url{https://link.aps.org/doi/10.1103/PhysRevLett.116.020601}
\BIBentrySTDinterwordspacing

\bibitem{philbin2016thermal}
\BIBentryALTinterwordspacing
T.~G. Philbin and J.~Anders, ``Thermal energies of classical and quantum damped oscillators coupled to reservoirs,'' \emph{Journal of Physics A: Mathematical and Theoretical}, vol.~49, no.~21, p. 215303, apr 2016. [Online]. Available: \url{https://dx.doi.org/10.1088/1751-8113/49/21/215303}
\BIBentrySTDinterwordspacing

\bibitem{jarzynski2004nonequilibrium}
\BIBentryALTinterwordspacing
C.~Jarzynski, ``Nonequilibrium work theorem for a system strongly coupled to a thermal environment,'' \emph{Journal of Statistical Mechanics: Theory and Experiment}, vol. 2004, no.~09, p. P09005, sep 2004. [Online]. Available: \url{https://dx.doi.org/10.1088/1742-5468/2004/09/P09005}
\BIBentrySTDinterwordspacing

\bibitem{campisi2009fluctuation}
\BIBentryALTinterwordspacing
M.~Campisi, P.~Talkner, and P.~H\"anggi, ``Fluctuation theorem for arbitrary open quantum systems,'' \emph{Phys. Rev. Lett.}, vol. 102, p. 210401, May 2009. [Online]. Available: \url{https://link.aps.org/doi/10.1103/PhysRevLett.102.210401}
\BIBentrySTDinterwordspacing

\bibitem{rivas2020strong}
\BIBentryALTinterwordspacing
A.~Rivas, ``Strong coupling thermodynamics of open quantum systems,'' \emph{Phys. Rev. Lett.}, vol. 124, p. 160601, Apr 2020. [Online]. Available: \url{https://link.aps.org/doi/10.1103/PhysRevLett.124.160601}
\BIBentrySTDinterwordspacing

\bibitem{talkner2020colloquium}
\BIBentryALTinterwordspacing
P.~Talkner and P.~H\"anggi, ``Colloquium: Statistical mechanics and thermodynamics at strong coupling: Quantum and classical,'' \emph{Rev. Mod. Phys.}, vol.~92, p. 041002, Oct 2020. [Online]. Available: \url{https://link.aps.org/doi/10.1103/RevModPhys.92.041002}
\BIBentrySTDinterwordspacing

\bibitem{strasberg2020measurability}
\BIBentryALTinterwordspacing
P.~Strasberg and M.~Esposito, ``Measurability of nonequilibrium thermodynamics in terms of the hamiltonian of mean force,'' \emph{Phys. Rev. E}, vol. 101, p. 050101, May 2020. [Online]. Available: \url{https://link.aps.org/doi/10.1103/PhysRevE.101.050101}
\BIBentrySTDinterwordspacing

\bibitem{cresser2021weak}
\BIBentryALTinterwordspacing
J.~D. Cresser and J.~Anders, ``Weak and ultrastrong coupling limits of the quantum mean force gibbs state,'' \emph{Phys. Rev. Lett.}, vol. 127, p. 250601, Dec 2021. [Online]. Available: \url{https://link.aps.org/doi/10.1103/PhysRevLett.127.250601}
\BIBentrySTDinterwordspacing

\bibitem{kumar24ultra}
\BIBentryALTinterwordspacing
P.~Kumar and S.~Ghosh, ``{Ultrastrong coupling limit to quantum mean force Gibbs state for anharmonic environment},'' \emph{The Journal of Chemical Physics}, vol. 161, no.~17, p. 174111, 11 2024. [Online]. Available: \url{https://doi.org/10.1063/5.0223734}
\BIBentrySTDinterwordspacing

\bibitem{PhysRevA.106.042209}
\BIBentryALTinterwordspacing
A.~Trushechkin, ``Quantum master equations and steady states for the ultrastrong-coupling limit and the strong-decoherence limit,'' \emph{Phys. Rev. A}, vol. 106, p. 042209, Oct 2022. [Online]. Available: \url{https://link.aps.org/doi/10.1103/PhysRevA.106.042209}
\BIBentrySTDinterwordspacing

\bibitem{jaksic1996on}
\BIBentryALTinterwordspacing
V.~Jakšić and C.-A. Pillet, ``On a model for quantum friction iii. ergodic properties of the spin-boson system,'' \emph{Communications in Mathematical Physics}, vol. 178, no.~3, pp. 627--651, 1996. [Online]. Available: \url{https://doi.org/10.1007/BF02108818}
\BIBentrySTDinterwordspacing

\bibitem{bach2000return}
\BIBentryALTinterwordspacing
V.~Bach, J.~Fröhlich, and I.~M. Sigal, ``Return to equilibrium,'' \emph{Journal of Mathematical Physics}, vol.~41, no.~6, pp. 3985--4060, 06 2000. [Online]. Available: \url{https://doi.org/10.1063/1.533334}
\BIBentrySTDinterwordspacing

\bibitem{merkli2007decoherence}
\BIBentryALTinterwordspacing
M.~Merkli, I.~M. Sigal, and G.~P. Berman, ``Decoherence and thermalization,'' \emph{Phys. Rev. Lett.}, vol.~98, p. 130401, Mar 2007. [Online]. Available: \url{https://link.aps.org/doi/10.1103/PhysRevLett.98.130401}
\BIBentrySTDinterwordspacing

\bibitem{merkli2008resonance}
\BIBentryALTinterwordspacing
M.~Merkli, I.~Sigal, and G.~Berman, ``Resonance theory of decoherence and thermalization,'' \emph{Annals of Physics}, vol. 323, no.~2, pp. 373--412, 2008. [Online]. Available: \url{https://www.sciencedirect.com/science/article/pii/S0003491607000589}
\BIBentrySTDinterwordspacing

\bibitem{merkli2008dynamics}
\BIBentryALTinterwordspacing
M.~Merkli, G.~Berman, and I.~Sigal, ``Dynamics of collective decoherence and thermalization,'' \emph{Annals of Physics}, vol. 323, no.~12, pp. 3091--3112, 2008. [Online]. Available: \url{https://www.sciencedirect.com/science/article/pii/S0003491608001140}
\BIBentrySTDinterwordspacing

\bibitem{merkli2001positive}
\BIBentryALTinterwordspacing
M.~Merkli, ``Positive commutators in non-equilibrium quantum statistical mechanics,'' \emph{Communications in Mathematical Physics}, vol. 223, no.~2, pp. 327--362, 2001. [Online]. Available: \url{https://doi.org/10.1007/s002200100545}
\BIBentrySTDinterwordspacing

\bibitem{frohlich2004another}
\BIBentryALTinterwordspacing
J.~Fröhlich and M.~Merkli, ``Another return of “return to equilibrium”,'' \emph{Communications in Mathematical Physics}, vol. 251, no.~2, pp. 235--262, 2004. [Online]. Available: \url{https://doi.org/10.1007/s00220-004-1176-6}
\BIBentrySTDinterwordspacing

\bibitem{merkli2022dynamics}
\BIBentryALTinterwordspacing
M.~Merkli, ``Dynamics of {O}pen {Q}uantum {S}ystems {I}, {O}scillation and {D}ecay,'' \emph{{Quantum}}, vol.~6, p. 615, Jan. 2022. [Online]. Available: \url{https://doi.org/10.22331/q-2022-01-03-615}
\BIBentrySTDinterwordspacing

\bibitem{merkli2022dynamicsb}
\BIBentryALTinterwordspacing
------, ``Dynamics of {O}pen {Q}uantum {S}ystems {II}, {M}arkovian {A}pproximation,'' \emph{{Quantum}}, vol.~6, p. 616, Jan. 2022. [Online]. Available: \url{https://doi.org/10.22331/q-2022-01-03-616}
\BIBentrySTDinterwordspacing

\bibitem{crowder2024invalidation}
\BIBentryALTinterwordspacing
E.~Crowder, L.~Lampert, G.~Manchanda, B.~Shoffeitt, S.~Gadamsetty, Y.~Pei, S.~Chaudhary, and D.~Davidovi\ifmmode~\acute{c}\else \'{c}\fi{}, ``Invalidation of the bloch-redfield equation in the sub-ohmic regime via a practical time-convolutionless fourth-order master equation,'' \emph{Phys. Rev. A}, vol. 109, p. 052205, May 2024. [Online]. Available: \url{https://link.aps.org/doi/10.1103/PhysRevA.109.052205}
\BIBentrySTDinterwordspacing

\bibitem{PhysRevE.110.014144}
\BIBentryALTinterwordspacing
M.~\L{}obejko, M.~Winczewski, G.~Su\'arez, R.~Alicki, and M.~Horodecki, ``Corrections to the hamiltonian induced by finite-strength coupling to the environment,'' \emph{Phys. Rev. E}, vol. 110, p. 014144, Jul 2024. [Online]. Available: \url{https://link.aps.org/doi/10.1103/PhysRevE.110.014144}
\BIBentrySTDinterwordspacing

\bibitem{CummulantMEVsBRME}
The Cumulant ME is a completely-positive ME, unlike BR or TCL-ME. Since a TCL2n-ME is a perturbative ME correct up to $O(\lambda^{2n})$, it can have issues with positivity up to this order. Here, we are not interested in this aspect of the TCL.

\bibitem{gilmore2006criteria}
\BIBentryALTinterwordspacing
J.~B. Gilmore and R.~H. McKenzie, ``Criteria for quantum coherent transfer of excitations between chromophores in a polar solvent,'' \emph{Chemical Physics Letters}, vol. 421, no.~1, pp. 266--271, 2006. [Online]. Available: \url{https://www.sciencedirect.com/science/article/pii/S0009261405019275}
\BIBentrySTDinterwordspacing

\bibitem{Gilmore_2005}
\BIBentryALTinterwordspacing
J.~Gilmore and R.~H. McKenzie, ``Spin boson models for quantum decoherence of electronic excitations of biomolecules and quantum dots in a solvent,'' \emph{Journal of Physics: Condensed Matter}, vol.~17, no.~10, p. 1735, feb 2005. [Online]. Available: \url{https://dx.doi.org/10.1088/0953-8984/17/10/028}
\BIBentrySTDinterwordspacing

\bibitem{gilmore2008quantum}
\BIBentryALTinterwordspacing
------, ``Quantum dynamics of electronic excitations in biomolecular chromophores: Role of the protein environment and solvent,'' \emph{The Journal of Physical Chemistry A}, vol. 112, no.~11, pp. 2162--2176, 2008, pMID: 18293949. [Online]. Available: \url{https://doi.org/10.1021/jp710243t}
\BIBentrySTDinterwordspacing

\bibitem{porras2008mesoscopic}
\BIBentryALTinterwordspacing
D.~Porras, F.~Marquardt, J.~von Delft, and J.~I. Cirac, ``Mesoscopic spin-boson models of trapped ions,'' \emph{Phys. Rev. A}, vol.~78, p. 010101, Jul 2008. [Online]. Available: \url{https://link.aps.org/doi/10.1103/PhysRevA.78.010101}
\BIBentrySTDinterwordspacing

\bibitem{cheche2001dynamics}
\BIBentryALTinterwordspacing
T.~O. Cheche and S.~H. Lin, ``Dynamics of the spin-boson hamiltonian by the projection operator technique: Applications to electron transfer reactions,'' \emph{Phys. Rev. E}, vol.~64, p. 061103, Nov 2001. [Online]. Available: \url{https://link.aps.org/doi/10.1103/PhysRevE.64.061103}
\BIBentrySTDinterwordspacing

\bibitem{merkli2013electron}
\BIBentryALTinterwordspacing
M.~Merkli, G.~P. Berman, and R.~Sayre, ``\BIBforeignlanguage{eng}{Electron transfer reactions: generalized spin-boson approach},'' \emph{\BIBforeignlanguage{eng}{Journal of mathematical chemistry}}, vol.~51, no.~3, p.~24, 2013. [Online]. Available: \url{https://doi.org/10.1007/s10910-012-0124-5}
\BIBentrySTDinterwordspacing

\bibitem{magazzu2018probing}
\BIBentryALTinterwordspacing
L.~Magazz{\`u}, P.~Forn-D{\'\i}az, R.~Belyansky, J.-L. Orgiazzi, M.~Yurtalan, M.~R. Otto, A.~Lupascu, C.~Wilson, and M.~Grifoni, ``Probing the strongly driven spin-boson model in a superconducting quantum circuit,'' \emph{Nature communications}, vol.~9, no.~1, p. 1403, 2018. [Online]. Available: \url{https://doi.org/10.1038/s41467-018-03626-w}
\BIBentrySTDinterwordspacing

\bibitem{purkayastha2020tunable}
\BIBentryALTinterwordspacing
A.~Purkayastha, G.~Guarnieri, M.~T. Mitchison, R.~Filip, and J.~Goold, ``Tunable phonon-induced steady-state coherence in a double-quantum-dot charge qubit,'' \emph{npj Quantum Information}, vol.~6, no.~1, p.~27, 2020. [Online]. Available: \url{https://doi.org/10.1038/s41534-020-0256-6}
\BIBentrySTDinterwordspacing

\bibitem{PhysRevLett.104.036801}
\BIBentryALTinterwordspacing
C.~Weber, A.~Fuhrer, C.~Fasth, G.~Lindwall, L.~Samuelson, and A.~Wacker, ``Probing confined phonon modes by transport through a nanowire double quantum dot,'' \emph{Phys. Rev. Lett.}, vol. 104, p. 036801, Jan 2010. [Online]. Available: \url{https://link.aps.org/doi/10.1103/PhysRevLett.104.036801}
\BIBentrySTDinterwordspacing

\bibitem{colless2014raman}
J.~Colless, X.~Croot, T.~M. Stace, A.~C. Doherty, S.~D. Barrett, H.~Lu, A.~Gossard, and D.~J. Reilly, ``Raman phonon emission in a driven double quantum dot,'' \emph{Nature communications}, vol.~5, no.~1, p. 3716, 2014.

\bibitem{PhysRevLett.120.097701}
\BIBentryALTinterwordspacing
T.~R. Hartke, Y.-Y. Liu, M.~J. Gullans, and J.~R. Petta, ``Microwave detection of electron-phonon interactions in a cavity-coupled double quantum dot,'' \emph{Phys. Rev. Lett.}, vol. 120, p. 097701, Feb 2018. [Online]. Available: \url{https://link.aps.org/doi/10.1103/PhysRevLett.120.097701}
\BIBentrySTDinterwordspacing

\bibitem{PhysRevB.97.035305}
\BIBentryALTinterwordspacing
M.~J. Gullans, J.~M. Taylor, and J.~R. Petta, ``Probing electron-phonon interactions in the charge-photon dynamics of cavity-coupled double quantum dots,'' \emph{Phys. Rev. B}, vol.~97, p. 035305, Jan 2018. [Online]. Available: \url{https://link.aps.org/doi/10.1103/PhysRevB.97.035305}
\BIBentrySTDinterwordspacing

\bibitem{PhysRevApplied.9.014030}
\BIBentryALTinterwordspacing
Y.-Y. Liu, J.~Stehlik, X.~Mi, T.~R. Hartke, M.~J. Gullans, and J.~R. Petta, ``On-chip quantum-dot light source for quantum-device readout,'' \emph{Phys. Rev. Appl.}, vol.~9, p. 014030, Jan 2018. [Online]. Available: \url{https://link.aps.org/doi/10.1103/PhysRevApplied.9.014030}
\BIBentrySTDinterwordspacing

\bibitem{PhysRevX.6.041027}
\BIBentryALTinterwordspacing
J.~Stehlik, Y.-Y. Liu, C.~Eichler, T.~R. Hartke, X.~Mi, M.~J. Gullans, J.~M. Taylor, and J.~R. Petta, ``Double quantum dot floquet gain medium,'' \emph{Phys. Rev. X}, vol.~6, p. 041027, Nov 2016. [Online]. Available: \url{https://link.aps.org/doi/10.1103/PhysRevX.6.041027}
\BIBentrySTDinterwordspacing

\bibitem{doi:10.1126/science.aaa2501}
\BIBentryALTinterwordspacing
Y.-Y. Liu, J.~Stehlik, C.~Eichler, M.~J. Gullans, J.~M. Taylor, and J.~R. Petta, ``Semiconductor double quantum dot micromaser,'' \emph{Science}, vol. 347, no. 6219, pp. 285--287, 2015. [Online]. Available: \url{https://www.science.org/doi/abs/10.1126/science.aaa2501}
\BIBentrySTDinterwordspacing

\bibitem{PhysRevA.57.120}
\BIBentryALTinterwordspacing
D.~Loss and D.~P. DiVincenzo, ``Quantum computation with quantum dots,'' \emph{Phys. Rev. A}, vol.~57, pp. 120--126, Jan 1998. [Online]. Available: \url{https://link.aps.org/doi/10.1103/PhysRevA.57.120}
\BIBentrySTDinterwordspacing

\bibitem{BRANDES2005315}
\BIBentryALTinterwordspacing
T.~Brandes, ``Coherent and collective quantum optical effects in mesoscopic systems,'' \emph{Physics Reports}, vol. 408, no.~5, pp. 315--474, 2005. [Online]. Available: \url{https://www.sciencedirect.com/science/article/pii/S0370157304005496}
\BIBentrySTDinterwordspacing

\bibitem{breuer2001time}
\BIBentryALTinterwordspacing
H.-P. Breuer, B.~Kappler, and F.~Petruccione, ``The time-convolutionless projection operator technique in the quantum theory of dissipation and decoherence,'' \emph{Annals of Physics}, vol. 291, no.~1, pp. 36--70, 2001. [Online]. Available: \url{https://www.sciencedirect.com/science/article/pii/S0003491601961524}
\BIBentrySTDinterwordspacing

\bibitem{TypoAndConventionDiff}
For $a_1=1/2$ and $a_3 = 0$, our expressions for the TCL2 and TCL4 coefficient integrals reduce to those \highlighttext{obtained by Breuer et al.} \cite{breuer2001time} except for a sign error in the latter for the expression corresponding to \cref{eqn_b34_triple_integral}. Comparing with \highlighttext{Breuer et al.}, also note a trivial difference of a factor of $1/2$ in \cref{eqn_a334_triple_integral} and \cref{eqn_b34_triple_integral}, coming from notational difference in the choice of vectorization (see \cref{eqn_vectorization_of_qubit_state}).

\bibitem{SupplementaryMaterial}
\highlighttext{See Supplemental Material at \url{https://arxiv.org/src/2411.08869v1/anc/AncillaryFileMasterEquationProject.nb} for a MATHEMATICA notebook with all the results of this work in electronic form.}

\bibitem{thomas1996calculus}
\BIBentryALTinterwordspacing
G.~Thomas, M.~Weir, and R.~Finney, \emph{\highlighttext{Calculus and Analytic Geometry}}, ser. Addison-Wesley world student series.\hskip 1em plus 0.5em minus 0.4em\relax Addison-Wesley, Boston, 1996. [Online]. Available: \url{https://books.google.co.in/books?id=GWIdAAAACAAJ}
\BIBentrySTDinterwordspacing

\bibitem{weber2003essential}
H.~J. Weber and G.~B. Arfken, \emph{Essential mathematical methods for physicists, ISE}.\hskip 1em plus 0.5em minus 0.4em\relax Elsevier, 2003.

\end{thebibliography}

\clearpage  

\appendix
\crefname{section}{Appendix}{Appendices}
\Crefname{section}{Appendix}{Appendices}
\makeatletter

\section{Some steady state and mean force Gibbs state results for generic open quantum system} \label{appendix_general_results}

Here, we summarize some results valid for a generic open-quantum system at arbitrary SE coupling. This will help simplify the accompanying TCL2 and TCL4 calculations and also identify some generic features in the SS and the TCL generator.

For a generic open quantum system, not necessarily for a qubit system or Bosonic environment or weak coupling, let the full SE Hamiltonian be given as
\begin{align}
    \hat{H}_{SE} &= \hat{H}_S + \hat{H}_E + \hat{H}_I,\\
    \hat{H}_I &= \lambda \hat{A} \otimes \hat{B},
\end{align}
where $\hat{A}$ and $\hat{B}$ are system and environment operators, respectively.

For a generic operator $\hat{O}$ in Schrodinger picture, the corresponding Heisenberg picture operator is defined as \cite{breuer2002theory}
\begin{align}
    \hat{O}_H(t) = e^{i t \hat{H}_{SE}} \hat{O} e^{-i t \hat{H}_{SE}},
\end{align}
and the corresponding equation of motion is given as 
\begin{align}
    \frac{d}{dt} \hat{O}_H(t) &= i [\hat{H}_{SE}, \hat{O}_H(t)] + \left(\frac{\partial \hat{O}}{\partial t}\right)_H.
\end{align}
Assuming that $\hat{O}$ is time independent (that is, in the Schrodinger picture, we have $\partial \hat{O}/\partial t = 0$). Then the equation of motion for $\hat{O}_H(t)$ simplifies as
\begin{align}
    \frac{d}{dt} \hat{O}_H(t) &= i [\hat{H}_{SE}, \hat{O}_H(t)],\\
    &= i e^{i t \hat{H}_{SE}} [\hat{H}_{SE},  \hat{O}] e^{-i t \hat{H}_{SE}}.\label{eqn_heisenberg_eqn_for_operators}
\end{align}
Next, let us have, for some system operator $\hat{O}_1$,
\begin{align}\label{eqn_o1_defn_appendix}
    [\hat{O}_1, \hat{A}] \equiv 0.
\end{align}
Then, in \cref{eqn_heisenberg_eqn_for_operators}, we have
\begin{align}
    \frac{d}{dt} \hat{O}_{1,H}
 (t) &= i e^{i t \hat{H}_{SE}} [\hat{H}_S,  \hat{O}_1] e^{-i t \hat{H}_{SE}},\\
    \implies \frac{d}{dt} \braket{\hat{O}_{1,H}(t)} &= \Tr \{ -i [\hat{O}_1, \hat{H}_S] \hat{\rho}_{SE}(t)\}.\label{eqn_o1_expec_evolve}
\end{align}
Hence, in the large time limit, we have
\begin{align}\label{eqn_o1_SS_derivative_a}
    \frac{d}{dt} \braket{\hat{O}_1}_{\hat{\rho}_S(t)} &= 0 = \lim_{t \to \infty} \Tr \{ -i [\hat{O}_1, \hat{H}_S] \hat{\rho}_{SE}(t)\}.
\end{align}
Defining
\begin{align}\label{eqn_O2_SS_0_a}
    \hat{O}_2 \equiv [\hat{O}_1, \hat{H}_S], 
\end{align}
for a generic open quantum system with a SS, we then have that,
\begin{align}\label{eqn_O2_SS_result}
    \braket{\hat{O}_2}_{SS} &= 0.
\end{align}

\subsection{Consistency with mean force Gibbs state}
We will now prove that for the MFGS at arbitrary SE coupling, we correspondingly have that
\begin{align}
    \braket{\hat{O}_2}_{MF} &= 0.
\end{align}
Proof: We have,
\begin{align}
    \braket{\hat{O}_2}_{MF} &\equiv \Tr_S\left\{ \hat{O}_2 \hat{\rho}_{MF}\right\},\\
    &= \Tr_S\left\{ \hat{O}_2 \sum_{n=0}^\infty Z^{-1} \Tr_E \exp\{-\beta \hat{H}_{SE}\}\right\},\\
    &= \Tr\left\{ \hat{O}_2 \sum_{n=0}^\infty \frac{1}{n! Z} (-\beta)^n \hat{H}_{SE}^n\right\},\\
    &= \sum_{n=0}^\infty \frac{1}{n! Z} (-\beta)^n \Tr\left\{ \hat{O}_2  \hat{H}_{SE}^n\right\},\\
    &= \sum_{n=0}^\infty \frac{1}{n! Z} (-\beta)^n \Tr\left\{ [\hat{O}_1, \hat{H}_S]  \hat{H}_{SE}^n\right\},
\end{align}
In \cref{appendix_MFGS_result}, we prove that, for any operators $\hat{O}, \hat{S}$ and $\hat{J}$, given that $[\hat{O},\hat{J}] = 0$, we have
\begin{align}
    \Tr \{ [\hat{O}, \hat{S}] (\hat{S} + \hat{J})^N  \} = 0.
\end{align}
where, $N$ is a whole number. Writing $\hat{O} \to \hat{O}_1$, $\hat{S} \to \hat{H}_S$ and $\hat{J} \to \hat{H}_E + \hat{H}_I$, we observe that we do have $[\hat{O}_1, \hat{H}_E + \hat{H}_I] = 0$. Hence, we get,
\begin{align}
    \Tr \{ [\hat{O}_1, \hat{H}_S] (\hat{H}_S + \hat{H}_E + \hat{H}_I)^N  \} &= 0,\\
    \implies \Tr\left\{ \hat{O}_2  \hat{H}_{SE}^n\right\} &= 0\\
    \implies \braket{\hat{O}_2}_{MF} &= 0.\label{eqn_O2_MFGS_result}
\end{align}

We note that for a $d$ dimensional system, there are generally $d$ orthogonal operators $\hat{O}_1$ that commute with $\hat{A}$, of which one is trivially the identity operator. Each of the remaining nontrivial operators $\hat{O}_1$ has a corresponding operator $\hat{O}_2$ for which the correspondence between MFGS and SS holds (\cref{eqn_O2_SS_result} and \cref{eqn_O2_MFGS_result}). Hence, although the general results on the limits on the SE coupling strength $\lambda$ and temperature $T$ for which the equivalence between SS and MFGS holds, are not available \cite{trushechkin2022quantum, jaksic1996on, bach2000return, merkli2007decoherence, merkli2008resonance, merkli2008dynamics, merkli2001positive, frohlich2004another, merkli2022dynamics, merkli2022dynamicsb}, for the expectation values of the class of operators, $\hat{O}_2$, at least, the equivalence exists for the entire range of SE parameters. For example, it can be verified that $\braket{\hat{O}_2}_{MF} = 0$ even for known analytical expression for the ultrastrong coupling MFGS \cite{cresser2021weak, kumar24ultra}.

\subsection{Proof of the commutation identity}\label{appendix_MFGS_result}

\begin{mybox}[Proposition]
Given
\begin{align}\label{eqn_O_and_J_commute}
    [\hat{O},\hat{J}] = 0,
\end{align}
to prove that
\begin{align} \label{eqn_statement_mfgs_commutator_identity}
    \Tr \{ [\hat{O}, \hat{S}] (\hat{S} + \hat{J})^N  \} = 0.
\end{align}
where $N$ is a whole number.
\end{mybox}

Proof: We have
\begin{align}\label{eqn_simplified_proof_statement_mfgs}
    \Tr \{ [\hat{O}, \hat{S}] (\hat{S} + \hat{J})^N  \} &\equiv T_1 - T_2,
\end{align}
where
\begin{align}
    T_1 &= \Tr\{\hat{O} \hat{S} (\hat{S} + \hat{J})^N\},\\
    T_2 &= \Tr\{\hat{S}\hat{O} (\hat{S} + \hat{J})^N\}.  
\end{align}

\highlighttext{Each of $T_1$ and $T_2$ will have $2^N$ total number of terms, with each term expressed as trace over product of $N + 2$ operators.}
The idea is that, for every term in $T_1$, \highlighttext{let us call it $e_1$}, we need to find a cyclically equivalent unique term in $T_2$, \highlighttext{which we will call $e_2$}, and which will cancel it under trace \highlighttext{(in \cref{eqn_simplified_proof_statement_mfgs}).

Let us symbolically represent the terms, $e_1$ and $e_2$, as
\begin{align}
    e_1 &= \Tr\{ \hat{O} \hat{S} X_1 ... X_{N} \},\\
    e_2 &= \Tr\{ \hat{S} \hat{O} Y_1 ... Y_{N} \}.
\end{align}
Here, $X_i, Y_i \in \{\hat{S}, \hat{J}\}$. Also, for mathematical convenience, let us define $X_0 = Y_0 = \hat{S}$. Then, the above expression transform as
\begin{align} \label{eqn_refined_e1}
    e_1 &= \Tr\{ \hat{O} X_0 X_1 ... X_{N} \},\\
    e_2 &= \Tr\{ Y_0 \hat{O} Y_1 ... Y_{N} \}. \label{eqn_refined_e2}
\end{align}


Then, for a given choice of $X_i$, let $0\leq k \leq N$ be the largest integer such that $X_k = \hat{S}$. Note that there will always be a solution for $k$ because $X_0 = \hat{S}$ (hence, $k=0$ is always a possible solution).

Next, by definition of $k$, for $n>k$, $X_n = \hat{J}$. Hence, $e_1$ has the form
\begin{align}
    e_1 &= \Tr\{\hat{O} X_0 ... X_{k-1} \hat{S} \hat{J} \hat{J} ... (N-k \text{ times})\}.
\end{align}
Using cyclic rotation under trace and the fact that $\hat{O}$ commutes with $\hat{J}$ (\cref{eqn_O_and_J_commute}), we get,
\begin{align}\label{eqn_e2_intuitive_form}
    e_1 &= \Tr\{ \hat{S} \hat{O} \hat{J} \hat{J} ... (N-k \text{ times}) X_0 ... X_{k-1}\},\\
    &= \Tr\{ X_k \hat{O} X_{k+1} ... X_N X_0 ... X_{k-1}\}, \label{eqn_before_going_to_e2}\\
    &\equiv \Tr\{ Y_0 \hat{O} Y_{1} ... Y_N\}, \label{eqn_after_going_to_e2}\\
    &= e_2. \label{eqn_e2_final}
\end{align}
That is, going from \cref{eqn_before_going_to_e2} to \cref{eqn_after_going_to_e2} \textit{defines} the map between $X_i$ and $Y_j$ that is required to go from $e_1$ to $e_2$. That is, we used the relation $Y_0 = X_k, Y_1 = X_{k+1}$ ... $Y_{N+1 -k} = X_0$... $Y_N = X_{k-1}$. More formally, this map is given by
\begin{align}\label{eqn_Y_and_X_map}
    Y_{n (\text{mod}(N+1))} &\equiv X_{(n+k) (\text{mod}(N+1))}.
\end{align}
We next need to prove that this map is bijective.
\cref{eqn_Y_and_X_map} is inherently a bijective function, but there is one nontriviality. The integer $k$ is defined given $e_1$ and we need a procedure to determine it given $e_2$. Once we know $k$ given $e_2$, then the corresponding $e_1$ can similarly be determined through `the reverse cyclic shift' as
\begin{align}\label{eqn_reverse_cyclic_shoft}
    X_{n (\text{mod}(N+1))} &\equiv Y_{(n-k) (\text{mod}(N+1))}.
\end{align}
We note that going from \cref{eqn_e2_intuitive_form} to \cref{eqn_e2_final} involves variable renaming and no further operator manipulation. Hence, \cref{eqn_e2_intuitive_form} is essentially the expression for $e_2$. To identify the value of $k$ from here, we just count the number of consecutive $\hat{J}$'s to the right of $\hat{O}$, which is numerically equal to $N-k$ (see \cref{eqn_e2_intuitive_form}). The value of $k$ can thus be determined, allowing us to determine the inverse map and proving the bijectivity.

\textbf{Summary: } We have proved that for every $e_1$, there exists a $e_2$ such that $e_1 = e_2$, and vice versa. In \cref{eqn_simplified_proof_statement_mfgs}, these two terms, $e_1$ and $e_2$, are going to cancel each other out, which proves our final result (\cref{eqn_statement_mfgs_commutator_identity}).}

\section{TCL2 integrals and their solution}\label{appendix_TCL_2_gen_and_sol}
The matrix element of the TCL2 generator $F^{(2)}(t)$ can be evaluated as \cite{breuer2001time, breuer2002theory}
\begin{align}\label{eqn_TCL2_F10}
F_{10}^{(2)}(t) = \int_0^t 4 a_1 a_3 \eta \left(t-t_1\right) \sin \left(\left(t-t_1\right) \Omega \right) \, dt_1,
\end{align}
\begin{align}
F_{11}^{(2)}(t) = \int_0^t -4 a_3^2 \nu \left(t-t_1\right) \, dt_1,
\end{align}
\begin{align}
F_{12}^{(2)}(t) = 0,
\end{align}
\begin{align}
F_{13}^{(2)}(t) = \int_0^t -4 a_1 a_3 \nu \left(t-t_1\right) \cos \left(\left(t-t_1\right) \Omega \right) \, dt_1,
\end{align}
\begin{align}
F_{20}^{(2)}(t) = \int_0^t -4 a_1 a_3 \eta \left(t-t_1\right) \left(\cos \left(\left(t-t_1\right) \Omega \right)-1\right) \, dt_1,
\end{align}
\begin{align}
F_{21}^{(2)}(t) = \int_0^t 4 a_1^2 \nu \left(t-t_1\right) \sin \left(\left(t-t_1\right) \Omega \right) \, dt_1,
\end{align}
\begin{align}
F_{22}^{(2)}(t) = \int_0^t -4 \nu \left(t-t_1\right) \left(a_1^2 \cos \left(\left(t-t_1\right) \Omega \right)+a_3^2\right) \, dt_1,
\end{align}
\begin{align}\label{eqn_TCL2_F23}
F_{23}^{(2)}(t) = \int_0^t -4 a_1 a_3 \nu \left(t-t_1\right) \sin \left(\left(t-t_1\right) \Omega \right) \, dt_1.
\end{align}
Here, we deliberately omitted some matrix elements of this $4 \times 4$ matrix. Trace preservation condition demands that $F_{0i}^{(2)}(t) = 0$. Also, the expression for the remaining matrix elements depends upon those already stated here, and are given as
\begin{align}\label{eqn_DQD_TCL2_gen_symmetry}
    F_{3k}^{(2)}(t) = \frac{a_1}{a_3} F_{1k}^{(2)}(t).
\end{align}
\cref{eqn_DQD_TCL2_gen_symmetry} can be understood from the argument given in \cref{sec_coh_correction}, valid for a generic open quantum system and not just for TCL-ME. Choosing $\hat{O}_1 = \hat{A}$ (\cref{eqn_o1_defn_appendix}), we have (\cref{eqn_o1_SS_derivative_a})
\begin{align}
    \frac{d}{dt} \braket{\hat{A}}_{\hat{\rho}_S(t)} &= \Tr \{ -i [\hat{A}, \hat{H}_S] \hat{\rho}_{S}(t)\}.
\end{align}
This means that the expectation value of $\hat{A}$ is only evolved by TCL0. Hence, for $n>1$, we have
\begin{align}
    \Tr\{ \hat{A} \mathcal{F}^{(2n)}(t) [\hat{\rho}]\} &= 0 \quad \forall \hat{\rho},\\
    \implies \braket{\hat{A} \mathcal{F}^{(2n)}(t) [\hat{\sigma}_k]} &= 0 \text{ replacing }\hat{\rho} \to \hat{\sigma}_k,\\
    \implies \braket{a_3 \hat{\sigma}_3 \mathcal{F}^{(2n)}(t) [\hat{\sigma}_k]} &= \braket{a_1 \hat{\sigma}_1 \mathcal{F}^{(2n)}(t) [\hat{\sigma}_k]},\\
    \implies a_3 F_{3k}^{(2n)}(t) &= a_1 F_{1k}^{(2n)}(t), \text{ using \cref{eqn_operator_to_matrix}}\\
    \implies F_{3k}^{(2)}(t) &= \frac{a_1}{a_3} F_{1k}^{(2)}(t).
\end{align}
This gives us the desired result (\cref{eqn_DQD_TCL2_gen_symmetry}), true for arbitrary order TCL-ME and not just TCL2.

Next, we solve these integrals assuming an arbitrary spectral density $J(\omega)$, and get a general expression for the TCL2 matrix elements as,
\begin{align}
F_{10}^{(2)} = -2 \pi  a_1 a_3 J(\Omega ),
\end{align}
\begin{align}
F_{11}^{(2)} = -2 \pi  a_3^2 f(0),
\end{align}
\begin{align}
F_{13}^{(2)} = -2 \pi  a_1 a_3 f(\Omega ),
\end{align}
\begin{align}
F_{20}^{(2)} = \int_0^{\infty } -\frac{4 a_1 a_3 \Omega ^2 J(\omega )}{\omega  \left(-\omega ^2+\Omega ^2\right)} \, d\omega,
\end{align}
\begin{align}
F_{21}^{(2)} = \int_0^{\infty } \frac{4 a_1^2 \Omega  f(\omega )}{-\omega ^2+\Omega ^2} \, d\omega,
\end{align}
\begin{align}
F_{22}^{(2)} = -2 \pi  a_1^2 f(\Omega )-2 \pi  a_3^2 f(0),
\end{align}
\begin{align}
F_{23}^{(2)} = \int_0^{\infty } -\frac{4 a_1 a_3 \Omega  f(\omega )}{-\omega ^2+\Omega ^2} \, d\omega.
\end{align}
Again, the remaining matrix elements are zero or can be obtained from these using \cref{eqn_DQD_TCL2_gen_symmetry}.

\section{TCL4 integral evaluation}\label{appendix_TCL4_integral}
We will now discuss in detail the evaluation of one of the TCL4 integrals, $F_{33}^{(4)}(t)$ (\cref{eqn_a334_triple_integral}) for a simpler case when $a_3 = 0$ and $a_1 = 1/2$. We first expand the integrand of \cref{eqn_a334_triple_integral} using the definition of $\nu(t)$ given in terms of an integral over $\omega$ (\cref{eqn_nut_expr}). Since we have assumed an antisymmetric spectral density, $J(\omega)$, this means that $f(\omega)$ (\cref{eqn_definition_of_f_omega}) is symmetric. Hence, we can extend the integration limit of $\omega$ from $(0,+ \infty)$ to $(-\infty, +\infty)$ and multiply a factor of $1/2$ to get the following expression for $F_{33}^{(4)}$,
\begin{widetext}
\begin{align}
    F_{33}^{(4)} = \lim_{t\to \infty} \frac{1}{4}\int_{-\infty}^\infty d \omega_2 &\int_{-\infty}^\infty d \omega_1 
    \int_0^{t} dt_1 \int_0^{t_1} dt_1  \int_0^{t_2} dt_3 
    f\left(\omega _1\right) f\left(\omega _2\right) \notag \\
    &\times \bigg\{\cos \left(\left(t-t_2\right) \omega _1\right) \cos \left(\left(t_1-t_3\right) \omega _2\right) \sin \left(\left(t_1-t_2\right) \Omega \right) \sin \left(\left(t-t_3\right) \Omega \right)\notag\\
    &+\cos \left(\left(t-t_3\right) \omega _1\right) \cos \left(\left(t_1-t_2\right) \omega _2\right) \sin \left(\left(t-t_2\right) \Omega \right) \sin \left(\left(t_1-t_3\right) \Omega \right)\bigg\}.
\end{align}
After analytically doing the triple time integral for arbitrary spectral density $J(\omega)$, which can be easily done using a computer algebra system like MATHEMATICA, we are then left with the remaining $\omega_1$ and $\omega_2$ integrals. The integrand now has the following form,
\begin{align}\label{eqn_result_triple_integral}
    F_{33}^{(4)} = \lim_{t\to \infty}  \int_{-\infty}^\infty d \omega_2 \int_{-\infty}^\infty d \omega_1 \sum_{n,m=-1}^{1} D_{n,m}(\omega_1, \omega_2) e^{i t (n \omega_1 + m \omega_2)},
\end{align}
where $D_{0,0}(\omega_1, \omega_2)  = 0$ while the other coefficients are non-trivial functions of $\omega_1$ and $\omega_2$.

Let us do the $\omega_1$ integral first. Then, in \cref{eqn_result_triple_integral}, for $n \neq 0$, the $\omega_1$ integral is straightforward to evaluate because for $n=\pm 1$, the large $t$ limit causes only the poles at the real line to contribute to the integral.
Hence, for $n \neq 0$ case, we are able to exactly evaluate the $\omega_1$ integral and eliminate the variable $\omega_1$ from the equation. The remaining $\omega_2$ integral is dealt with likewise, giving out the final expression either as a scalar or in terms of an integral over $\omega_2$. After renaming $\omega_2 \to \omega$, the integral is evaluated as,
\begin{align}
    \lim_{t\to \infty} \int_{-\infty}^\infty d \omega_2 \int_{-\infty}^\infty d \omega_1 \sum_{n = \pm 1} \sum_{m=-1}^{1} D_{n,m}(\omega_1, \omega_2) e^{i t (n \omega_1 + m \omega_2)} = \frac{1}{2}\int_{-\infty}^\infty d \omega \frac{\pi  \Omega  f(\omega ) f'(\Omega )}{4 \omega ^2-4 \Omega ^2}.
\end{align}
Although it does not matter in this case, the contour for the $\omega$ integral above is defined to pass any pole at the real line from above (see \cref{fig_contour_avoiding_poles}). We note that we can always achieve this by adding the contribution of the real poles to the integral explicitly. This is done to make the final result simpler in form because such a contour naturally arises for the $n=0$ case that we will discuss next.
\begin{figure}[htbp]
    \centering
    \begin{tikzpicture}[scale=1,decoration={markings,mark=at position 0.6 with {\arrow{>}}}] 
        \draw[->] (-5,0) -- (5,0) node[right] {\(\text{Re}(\omega_1)\)};
        \draw[->] (0,-1) -- (0,2.5) node[above] {\(\text{Im}(\omega_1)\)};
        \draw[thick,postaction={decorate}] (-4,0) -- (4,0); 

        \draw[thick, blue, postaction={decorate}] 
        plot [smooth, tension=1] coordinates {(4,0) (2,0.5) (0,0.5) (-2,1) (-4,0)};

        \node[blue, above right] at (-2,1.2) {$C$};
        
        \fill[red] (1,1) circle (2pt);
        \node[red, above right] at (1,1.2) {$P$};

        \fill[blue] (2,0) circle (2pt);
        \fill[blue] (-2,0) circle (2pt);
        \fill[blue] (0,0) circle (2pt);

    \end{tikzpicture}
    \caption{Completing the contour for $\omega_1$ integral from $(-\infty, \infty)$. }
    \label{fig_quarter_contour_above}
\end{figure}
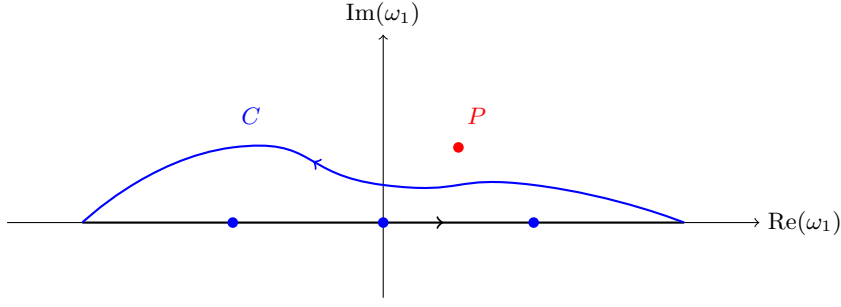

The $n=0$ case is more complicated, as the $\omega_1$ integral cannot be readily done for a general $f(\omega)$. In this case, we assume a contour $C$ as shown in \cref{fig_quarter_contour_above}, which is arbitrarily chosen in the positive imaginary plane such that the corresponding integral is non-divergent, though not necessarily zero. For simplicity of the discussion, we also assume that the contour $C$ is chosen such that it does not contain any extra poles of the integrand as a function of $\omega_1$ (for example, note how, in \cref{fig_quarter_contour_above}, the contour $C$ does not include the pole $P$). Then, we have
\begin{align}\label{eqn_n_equals_0_case_F334}
    \int_{-\infty}^\infty d \omega_1 \sum_{m= \pm 1} D_{0,m}(\omega_1, \omega_2) e^{i t m \omega_2} &= \sum_{m= \pm 1} \left(\pi i R_{m}(\omega_2) e^{i t m \omega_2} - \int_{C}  D_{0,m}(z, \omega_2) e^{i t m \omega_2}\, dz \right).
\end{align}
Here, $R_{m}(\omega_2)$ is, for a fixed value of $\omega_2$, the sum of the residues of the poles of the function $D_{0,m}(\omega_1, \omega_2)$ at real values of $\omega_1$ (see \cref{fig_quarter_contour_above}). Here, the expression for the sum over residues, $R_{m}(\omega_2)$, is expressed symbolically because we have not assumed any functional form of the spectral density, $J(\omega)$.

Next, since $D_{0,0} = 0$, as mentioned before, the remaining $\omega_2$ integral can be evaluated analytically because in the large $t$ limit, only the poles at the real value of $\omega_2$ will contribute. Once the $\omega_2$ variable is eliminated in this manner, the integrand becomes a function of $\omega_1$ in the form of a finite scalar and as an integral over the contour $C$.

In more detail, let $\omega_2$ equal to $c_i$ and $d_i$ be the real values at which the function $R_{m}(\omega_2)$ and $D_{0,m}(z, \omega_2)$ have poles, where the corresponding residue functions are symbolically denoted as $\Bar{R}_{m}(c_i)$ and $\Bar{D}_{0,m}(z, d_i)$, respectively. Then, for $m=1$ in \cref{eqn_n_equals_0_case_F334}, for example, we have,
\begin{align}\label{eqn_result_n_equals_zero_integral}
     \int_{-\infty}^\infty d \omega_2 \int_{-\infty}^\infty d \omega_1  D_{0,1}(\omega_1, \omega_2) e^{i t \omega_2} &= \pi i \sum_i \left( \pi i \Bar{R}_{1}(c_i) e^{i t c_i} - \int_{C} \Bar{D}_{0,1}(z, d_i) e^{i t d_i}\, dz \right).
\end{align}
We note that the only integral now left is over a contour $C$ (\cref{fig_quarter_contour_above}) which avoids any poles on the real line by passing it from above (see \cref{fig_contour_avoiding_poles}). Keeping this in mind, after renaming $\omega_1 \to \omega$, the final result comes out as
\begin{align}
    \int_0^\infty d \omega_2 \int_0^\infty d \omega_1 \sum_{m= \pm 1} D_{0,m}(\omega_1, \omega_2) e^{i t m \omega_2} &= \frac{1}{2}\int_{-\infty}^\infty d \omega \frac{\pi  \Omega ^2 f(\omega ) f(\Omega )}{2 \left(\omega ^2-\Omega ^2\right)^2}.
\end{align}
Combined, we have,
\begin{align}
    F_{33}^{(4)} &= \frac{\pi  \Omega}{4} \int_{-\infty}^\infty d\omega   \left(\frac{f(\omega ) f'(\Omega )}{2 \omega ^2-2 \Omega ^2}+\frac{\Omega  f(\omega ) f(\Omega )}{\left(\omega ^2-\Omega ^2\right)^2}\right)
\end{align}
which agrees with \cref{eqn_f334_result} for $a_3 = 0$ and $a_1=1/2$. The expression of $F_{33}^{(4)}$ for general values of $a_1$ and $a_3$ and also the corresponding expression for $F_{30}^{(4)}$ can similarly be evaluated using exactly the same steps mentioned here.
\end{widetext}
\subsection{Dependence of the integral on the order of the integration}\label{appendix_fubini_effect}

For the $n=0$ case above, we took a complicated route to evaluate the integral because it is not possible to readily evaluate the $\omega_1$ integral for general $f(\omega_1)$. We note that since $D_{00}(\omega_1, \omega_2) = 0$, the corresponding $\omega_2$ integral \textit{can} be evaluated explicitly at this point, and hence, one may get tempted to simplify the calculations by doing the $\omega_2$ integral first. Yet, this would lead to discrepancy, as here we show that the \highlighttext{chosen} order of integration at this point has to be respected \highlighttext{(i.e., integrating $\omega_1$ first).}

\subsubsection{Illustrative example}
\highlighttext{Before getting into the detail of this specific situation, let us illustrate the nature of the discrepancy through an example. Let us define
\begin{align}
    Q_1(f) &\equiv \int_{-\infty}^\infty dx \int_{-\infty}^\infty dy f(x,y)\\
    Q_2(f) &\equiv \int_{-\infty}^\infty dy \int_{-\infty}^\infty dx f(x,y)\\
    S(f) &\equiv Q_1(f) - Q_2(f)
\end{align}
Here, $Q_i$ and $S$ are functionals of $f$. Next, let us say we are interested in evaluating the integral
\begin{align}
    I &\equiv \int_{-\infty}^\infty dx \int_{-\infty}^\infty dy \left[ g(x,y) + h(x,y) \right]\\
    &= Q_1(g + h)
\end{align}
where, the functions $g(x,y)$ and $h(x,y)$ are \textit{defined} such that we have 
\begin{align}\label{eqn_s_opposite_sign}
    S(g) &\equiv -S(h) \neq 0
\end{align}
That is, they both individually violate Fubini's theorem \cite{thomas1996calculus} but under the constraint given in \cref{eqn_s_opposite_sign}. Then, for the sum of these two functions (i.e., $g(x,y) + h(x,y)$), the discrepancy disappears as following
\begin{align}
    S(g + h) &= S(g) + S(h) = 0.
\end{align}
Moreover, replacing the definition of $S$ in \cref{eqn_s_opposite_sign}, we get
\begin{align}
    Q_1(g) &- Q_2(g) = - Q_1(h) + Q_2(h)\\
    \implies Q_1(g) &+ Q_1(h) = Q_2(h) + Q_2(g)\\
    \implies I &= Q_1(g + h) = Q_2(g + h)
\end{align}
That is, the order in which the integral is carried out does not matter as long as it is done in the same order for both the individual terms $g$ and $h$.

Next, also note that if we are not consistent with the order of integration for both the individual terms, then we will encounter a discrepancy as following
\begin{align}
    Q_1(h) + Q_2(g) \neq Q_1(h + g).
\end{align}
In the case of the TCL4 integral, we encounter something similar.}
\begin{widetext}
\subsubsection{Specific details}

For $n=0$ and $m=1$ in the original integrand (\cref{eqn_result_triple_integral}), for example, $D_{0,1}(\omega_1, \omega_2) e^{i t \omega_2}$ has the precise form given as
\begin{align}\label{eqn_problematic_fubini_term}
    D_{0,1}(\omega_1, \omega_2) e^{i t \omega_2} &= \frac{\omega _2 \Omega  f\left(\omega _1\right) f\left(\omega _2\right) e^{i t \omega _2} \left(-\omega _2 \sin (t \Omega )-i \Omega  \cos (t \Omega )\right)}{2 \left(\omega _1^2-\omega _2^2\right) \left(\Omega ^2-\omega _1^2\right) \left(\Omega ^2-\omega _2^2\right)}. 
\end{align}
Note that since $f(\omega)$ is a symmetric function, the integrand is symmetric in $\omega_1$ but not so in $\omega_2$.
We have already discussed how to proceed if we do the $\omega_1$ integral first using the contour C represented in \cref{fig_quarter_contour_above}. We note that due to symmetry, the poles for $\omega_1$ at $\pm \omega_2$ and $\pm \Omega$ do not contribute to the integral ($R_{m}(\omega_2) = 0$ in \cref{eqn_n_equals_0_case_F334}). Hence, the $\omega_1$ integral is transformed as just an integral over the contour $C$ (\cref{fig_quarter_contour_above}). Afterwards, when we do the $\omega_2$ integral, in large $t$ limit, only the poles at the real values of $\omega_2$ will contribute to the integral. The only real poles that remain are for $\omega_2 = \pm \Omega$. This is because on the contour $C$, $\omega_1$ now attains a positive imaginary value, and hence the poles at $\omega_2 = \pm \omega_1$ have been lifted from the real axis of $\omega_2$.

On the other hand, if we try the $\omega_2$ integral first, then the real poles at $\omega_2 = \pm \Omega$ will give the exact same result as the integral before. But unlike in the last case, there will be some extra terms because of the poles at $\omega_2 = \pm \omega_1$ that were not there earlier. In fact, these extra terms give divergent contributions in the large $t$ limit, which is the limit we are interested in.

For this reason, we need to be consistent with the choice of the order of integration over the variables $\omega_1$ and $\omega_2$ for all the terms in the original integrand (\cref{eqn_result_triple_integral}). In other words, it does not matter whether we integrate $\omega_1$ first or $\omega_2$, but we need to be consistent with this choice for all terms.

Now, we will explicitly calculate the extra divergent term that we get when we do the $\omega_2$ integral first in \cref{eqn_problematic_fubini_term}. This comes from the $\omega_2 = \pm \omega_1$ poles, which contribute
\begin{align}
    \text{Extra Term} &= \pi i \int_{-\infty}^\infty d\omega_1 \left(\frac{i \Omega  f\left(\omega _1\right){}^2 e^{-i t \omega _1} \left(\Omega  \cos (t \Omega )+i \omega _1 \sin (t \Omega )\right)}{4 \left(\Omega ^2-\omega _1^2\right){}^2}+\frac{\Omega  f\left(\omega _1\right){}^2 e^{i t \omega _1} \left(\omega _1 \sin (t \Omega )+i \Omega  \cos (t \Omega )\right)}{4 \left(\Omega ^2-\omega _1^2\right){}^2}\right)
\end{align}
The $\omega_1$ contour for $e^{-i t \omega _1}$ and $e^{i t \omega _1}$ will be completed from below and above, respectively, with only the poles at $\omega_1 = \pm \Omega$ at the real line contributing to the $\omega_1$ integral in the large $t$ limit. This gives
\begin{align}\label{eqn_final_extra_term}
    \text{Extra Term} &= -\frac{\pi ^2 f(\Omega )^2 (\sin (2 t \Omega )-2 t \Omega )}{8 \Omega }
\end{align}
\end{widetext}
which, as we claimed, is divergent in the large $t$ limit. If we do the $\omega_2$ integration first, followed by $\omega_1$ integration, for all the 8 terms in the original integrand (\cref{eqn_result_triple_integral}), the sum of the extra terms (those corresponding to \cref{eqn_final_extra_term}) will vanish.

\cref{eqn_final_extra_term} corresponds to the extra term coming from the $(n=0, m=1)$ term in the original integrand (\cref{eqn_result_triple_integral}). The contribution from the corresponding $(n=0, m=-1)$ term will be exactly the same, while those from $(n=\pm 1, m=0)$ terms will be of the same magnitude but opposite sign, hence leading to cancellation. The remaining terms in \cref{eqn_result_triple_integral} do not lead to any extra term.

\section{TCL2 and TCL4 calculations for Ohmic spectral density with Drude cutoff}\label{appendix_TCL2_and_TCL4_for_Drude_cutoff}

Next, we discuss the calculations for TCL2 and TCL4 generators ($F^{(2)}$ and $F^{(4)}$, respectively) for Ohmic spectral density with Drude cutoff, defined to be of the form
\begin{align}
    J_D(\omega) = \frac{\gamma  \Lambda ^2 \omega }{\Lambda ^2+\omega ^2},
\end{align}
where $\gamma$ is the SE coupling parameter and $\Lambda$ is the environment cutoff frequency.

The two-point correlation functions, $\nu(t)$ and $\eta(t)$ (\cref{eqn_eta_expr} and \cref{eqn_nut_expr}) can be exactly calculated for $J_D(\omega)$ using contour integral as \cite{breuer2002theory}
\begin{align}
    \eta_D(t) &= -\frac{1}{2} \pi  \gamma  \Lambda ^2 e^{-t \Lambda},\\
    \nu_D(t) &= \sum_{n = -\infty}^\infty \frac{\pi  \gamma  \Lambda ^2 \left(\Lambda  e^{-\Lambda t}-\frac{2 \pi  | n|  e^{-\frac{2 \pi  t | n| }{\beta }}}{\beta }\right)}{\beta  \left(\Lambda ^2-\frac{4 \pi ^2 | n| ^2}{\beta ^2}\right)}.
\end{align}
Hence, these two-point correlation functions can be used directly to evaluate the time integrals of TCL2 and TCL4 coefficients (\cref{eqn_TCL2_F10}-\cref{eqn_TCL2_F23} and \cref{eqn_a334_triple_integral} and \cref{eqn_b34_triple_integral}) in the long time limit. This avoids the need to do the double $\omega$ integrals that were done in the case of our general results. The final result is then expressed in terms of a scalar, a single, or double infinite summation, which can be analytically evaluated in closed form or approximated numerically.

To simplify the calculations involved, we set $a_3 = 0$ and $a_1 = 1$ in the SBM Hamiltonian (\cref{eqn_spin_boson_A}). Then the non-zero coefficients for the TCL2 generator are given as
\begin{align}
F_{21}^{(2)} &= \frac{2 \gamma  \Lambda  \Omega}{\beta  \left(\Lambda ^2+\Omega ^2\right)}  
\bigg(2 \pi +\beta  \Lambda  \bigg[-2 \psi ^{(0)}\left(\frac{\beta  \Lambda }{2 \pi }+1\right)+\notag\\
&\quad \quad \psi ^{(0)}\left(1-\frac{i \beta  \Omega }{2 \pi }\right)+\psi ^{(0)}\left(\frac{i \beta  \Omega }{2 \pi }+1\right)\bigg]\bigg),
\end{align}
\begin{align}
F_{22}^{(2)} = -\frac{2 \pi  \gamma  \Lambda ^2 \Omega  \coth \left(\frac{\beta  \Omega }{2}\right)}{\Lambda ^2+\Omega ^2},
\end{align}
\begin{align}
F_{30}^{(2)} = -\frac{2 \pi  \gamma  \Lambda ^2 \Omega }{\Lambda ^2+\Omega ^2},
\end{align}
\begin{align}
F_{33}^{(2)} = -\frac{2 \pi  \gamma  \Lambda ^2 \Omega  \coth \left(\frac{\beta  \Omega }{2}\right)}{\Lambda ^2+\Omega ^2}.
\end{align}
Here and in subsequent equations, $\psi ^{(0)}(x)$ and $\psi ^{(1)}(x)$ are the digamma and trigamma functions, respectively \cite{weber2003essential}. The TCL4 coefficients $F_{33}^{(4)}$ and $F_{30}^{(4)}$ can also be evaluated as
\begin{widetext}
\begin{align}
    F_{30}^{(4)} &= \frac{2 \gamma ^2 \Lambda ^3 \Omega}{\beta  \left(\Lambda
   ^2+\Omega ^2\right)^3}  \bigg\{4 \pi  \beta  \Lambda  \left(\Lambda
   ^2-2 \Omega ^2\right) \psi ^{(0)}\left(\frac{\beta  \Lambda }{2 \pi
   }+1\right) +8 \pi ^2 \Omega ^2-\beta  \Lambda  \bigg[2 \pi  \left(\Lambda ^2-i \Lambda  \Omega
   -2 \Omega ^2\right) \psi ^{(0)}\left(1-\frac{i \beta  \Omega }{2 \pi
   }\right)\notag\\
   &+ 2 \pi  \left(\Lambda ^2+i \Lambda  \Omega -2 \Omega ^2\right) \psi
   ^{(0)}\left(\frac{i \beta  \Omega }{2 \pi }+1\right)-i \beta  \Omega 
   \left(\Lambda ^2+\Omega ^2\right) \left(\psi ^{(1)}\left(1-\frac{i \beta 
   \Omega }{2 \pi }\right)-\psi ^{(1)}\left(\frac{i \beta  \Omega }{2 \pi
   }+1\right)\right)\bigg]\bigg\},
\end{align}
\begin{align}
F_{33}^{(4)} &= \frac{\gamma ^2 \Lambda ^3 \Omega }{\pi  \beta ^2 \left(\Lambda ^2+\Omega ^2\right)^3}\bigg\{-2 i \pi  \beta ^2 \Lambda  \left(\Lambda ^2-3 \Omega ^2\right) \psi ^{(0)}\left(-\frac{i \beta  \Omega }{2 \pi }\right)^2+2 i \pi  \beta ^2 \Lambda  \left(\Lambda ^2-3 \Omega ^2\right) \psi ^{(0)}\left(\frac{i \beta  \Omega }{2 \pi }\right)^2+32 \pi ^3 \Omega \notag\\
&\quad \quad +2 \beta  \psi ^{(0)}\left(-\frac{i \beta  \Omega }{2 \pi }\right) \left(2 i \pi ^2 (\Lambda +i \Omega ) (\Lambda +3 i \Omega )-\beta ^2 \Lambda  \Omega  \left(\Lambda ^2+\Omega ^2\right) \psi ^{(1)}\left(-\frac{i \beta  \Omega }{2 \pi }\right)\right) \notag\\
&\quad \quad -2 i \pi  \beta ^2 (\Lambda -i \Omega ) (\Lambda +i \Omega ) \left((\Lambda +i \Omega ) \psi ^{(1)}\left(-\frac{i \beta  \Omega }{2 \pi }\right)-(\Lambda -i \Omega ) \psi ^{(1)}\left(\frac{i \beta  \Omega }{2 \pi }\right)\right) \notag\\
&\quad \quad +2 \beta  \psi ^{(0)}\left(\frac{i \beta  \Omega }{2 \pi }\right) \left(-\Lambda \beta ^2 \Omega  \left(\Lambda ^2+\Omega ^2\right) \psi ^{(1)}\left(\frac{i \beta  \Omega }{2 \pi }\right)-2 i \pi ^2 (\Lambda -i \Omega ) (\Lambda -3 i \Omega )\right) \notag\\
&\quad \quad -2 \pi ^2 \beta ^2 \Lambda  \psi ^{(0)}\left(\frac{\beta  \Lambda }{2 \pi }\right) \text{csch}^2\left(\frac{\beta  \Omega }{2}\right) \left(\beta  \Omega  \left(\Lambda ^2+\Omega ^2\right)-\left(\Lambda ^2-3 \Omega ^2\right) \sinh (\beta  \Omega )\right)\bigg\}.
\end{align}
\end{widetext}

We are also able to calculate the TCL2 and TCL4 matrix elements for the general SBM (for $a_3\neq0$), but are unable to easily reduce one of the infinite sums for $F_{33}^{(4)}$ in closed analytical form. However, we are able to approximately do the summation numerically and show its consistency with our main results.

\end{document}